\documentclass[12pt,a4paper]{article}

\usepackage{epsfig,graphicx}
\usepackage{amsmath,amsfonts,amssymb}
\usepackage{citesort}

\newcommand{\unit}{\leavevmode\hbox{\small1\kern-3.6pt\normalsize1}}

\def\neumass{m_{\tilde\chi_1^0}}
\newcommand{\crosssec}{\sigma_{\tilde\chi^0_1-p}}

\def\neut{\tilde\chi_1^0}

\def\asusy{a^{\rm SUSY}_\mu}

\def\lsim{\raise0.3ex\hbox{$\;<$\kern-0.75em\raise-1.1ex\hbox{$\sim\;$}}}
\def\gsim{\raise0.3ex\hbox{$\;>$\kern-0.75em\raise-1.1ex\hbox{$\sim\;$}}}
\def\nmh{{\sc nmhdecay}}

\newcommand{\captions}{\sf\caption}

\def\relic{\Omega_{\neut}\,h^2}

\begin{document}
  
\thispagestyle{empty}
\begin{flushright}
  FTUAM 07/01\\
  IFT-UAM/CSIC-07-02\\
  HIP-2006-02/TH\\
  LPT-ORSAY-07-02\\
  hep-ph/0701271\\
  \vspace*{5mm}{31 January 2007}
\end{flushright}

\begin{center}
  {\Large \textbf{Phenomenological viability of neutralino dark matter
      in the NMSSM} }
  
  \vspace{0.5cm}
  D.\,G.~Cerde\~no\,$^{a,b}$, E. Gabrielli\,$^{c}$,
  D.\,E.~L\'opez-Fogliani\,$^{a,b}$, C.~Mu\~noz\,$^{a,b}$,
  A.\,M.~Teixeira\,$^{d}$\\[0.2cm] 
    
  {$^{a}$\textit{Departamento de F\'{\i}sica Te\'{o}rica C-XI,
      Universidad Aut\'{o}noma de Madrid,\\[0pt] Cantoblanco, E-28049
      Madrid, Spain}}\\[0pt] 
  {$^{b}$\textit{Instituto de F\'{\i}sica Te\'{o}rica C-XVI, Universidad
      Aut\'{o}noma de Madrid,\\[0pt] Cantoblanco, E-28049 Madrid,
      Spain}}\\[0pt] 
  {$^{c}$\textit{Helsinki Institute of Physics P.O.B 64, 00014
      University of Helsinki, Finland}} \\[0pt] 
  {$^{d}$\textit {Laboratoire de Physique Th\'eorique, Universit\'e de
      Paris-Sud XI, B\^atiment 201, F-91405 Orsay Cedex, France}}\\[0pt]

  \begin{abstract}
    The viability of the lightest neutralino as a
    dark matter candidate in the Next-to-Minimal Supersymmetric
    Standard Model is analysed. 
    We carry out a thorough analysis of the
    parameter space,
    taking into account accelerator
    constraints as well as bounds on low-energy observables, such as
    the muon
    anomalous magnetic moment and rare $K$ and $B$ meson decays.
    The neutralino relic density is also evaluated and consistency
    with present bounds imposed.
    Finally, the neutralino direct 
    detection cross section is calculated in
    the allowed regions of the parameter space and compared to the
    sensitivities of present and projected dark matter experiments. 
    Regions of the parameter space are found where experimental
    constraints are fulfilled, the lightest
    neutralino has the correct relic abundance and its detection cross
    section is within the reach of dark matter detectors. This is
    possible in the presence of very light singlet-like Higgses and 
    when the neutralino is either light enough so that some
    annihilation channels are kinematically forbidden, or has a large
    singlino component.
  \end{abstract}
\end{center}

\vspace*{5mm}\hspace*{3mm}
	\newpage

\section{Introduction}\label{intro}

One of the most interesting candidates for dark matter, within the
class of Weakly Interactive Massive Particles 
(WIMPs), is the lightest neutralino ($\tilde \chi^0_1$), 
which arises in the context of
R-parity conserving supersymmetric (SUSY) models \cite{old}. 
Although neutralino dark matter has been extensively examined in
the framework of the Minimal Supersymmetric Standard Model (MSSM) 
\cite{lightreview}, this is not the case of 
the Next-to-Minimal Supersymmetric Standard Model (NMSSM)
\cite{NMSSM1}.

The NMSSM is an extension of the MSSM
by a singlet superfield $S$, which provides an elegant solution to the 
so-called $\mu$ problem of the MSSM \cite{mupb}. 
In this case, an effective $\mu$ term is
dynamically generated when the scalar component of $S$
acquires a vacuum expectation value (VEV) of order of the SUSY breaking 
scale. In addition, and when compared to the MSSM, the NMSSM renders
less severe the ``little fine tuning problem'' of the
MSSM~\cite{Bast}, or equivalently, the non-observation of a neutral
CP-even Higgs boson at LEP II. The presence of additional fields,
namely an extra CP-even and CP-odd neutral Higgs bosons, as well as a
fifth neutralino, leads to a richer and more complex 
phenomenology~\cite{NMSSM2,NLEP,NLHC,NHIGGS,NMHDECAY}. 
In particular, a very light neutralino may
be present~\cite{NLEP}. The upper bound on the mass of the lightest
Higgs state is larger than in the MSSM~\cite{NHIGGS}. Moreover, a very 
light Higgs boson is not experimentally excluded.
All these properties also contribute to the emergence
of dark matter scenarios that can be very different from those
encountered in the MSSM, both regarding the relic density and the
prospects for direct detection.

The NMSSM must comply with a large number of experimental 
constraints. In addition to LEP II and Tevatron limits on
the spectrum, one should also take into account 
SUSY contributions to low-energy observables.
The most stringent bounds arise 
from kaon and $B$ decays, as well as from the 
muon anomalous magnetic moment, $\asusy$.
Naturally, in order to be a viable dark matter candidate,
the NMSSM lightest neutralino must also satisfy
the present astrophysical bounds on the relic abundance. 
It is important to notice that, 
in addition to the channels already present in the MSSM, the NMSSM
offers new kinematically viable annihilation
channels and potential
resonances~\cite{Nrelden,Belanger:2005kh,Gunion:2005rw}, due to the 
presence of light scalar and pseudoscalar Higgses.

The direct detection of neutralino dark
matter has been discussed in the framework of the NMSSM in
Refs.~\cite{Ndirdet,bk,Cerdeno:2004xw,Gunion:2005rw}.
In particular, and as pointed 
out in~\cite{Cerdeno:2004xw}, the exchange of very light Higgses 
can lead to large direct detection cross sections,
within the reach of the present generation of dark matter
detectors. 
There are currently a large number of facilities devoted to 
the direct detection of WIMPs through their elastic scattering with nuclei.
Among them, it is important to refer to the DAMA 
collaboration~\cite{halo,experimento1} (which has already reported
data favouring the existence of a WIMP signal\footnote{Notice that the
DAMA result has not been confirmed by the other experiments. 
For attempts to show that DAMA and these experiments
might not be in conflict, see Ref.~\cite{conflict}.}), 
CDMS Soudan~\cite{soudan}, EDELWEISS~\cite{edelweiss} and 
ZEPLIN I~\cite{ZEPLINI}, as well as the upcoming detectors working
with 1 tonne of Ge/Xe~\cite{xenon}.

The purpose of the present work is to extend the analysis
of Ref.~\cite{Cerdeno:2004xw}. In addition to upgrading the
experimental constraints already present in the previous work, we now
include the computation of the SUSY contribution to the 
muon anomalous magnetic moment, 
and bounds from $K$- and $B$-meson decays. 
Moreover, we incorporate
the evaluation of the relic abundance of the lightest neutralino in
the NMSSM.

Our work is organised as follows.
We briefly address the most relevant aspects of the NMSSM in
Section\,\ref{NMSSM}, paying especial attention to the minimisation
of the Higgs potential and to the new features of the Higgs and neutralino
sectors. We also comment on the various constraints on the low-energy
observables that will be included in the analysis.
In Section\,\ref{DM:NMSSM} we introduce the NMSSM lightest neutralino
as a dark matter candidate and discuss the potential implications 
regarding its detection cross section and relic abundance. 
Our results are presented in 
Section\,\ref{discussion}, where we first analyse the effect of the 
experimental
constraints on low-energy observables, $\asusy$ and $b\to s\gamma$, on
the parameter space (Sec.\,\ref{constraints:up}). 
We then include the constraint on the
neutralino relic density (Sec.\,\ref{constraints:om}). 
Finally, taking all these bounds into
account, we evaluate the neutralino detection cross section on the
allowed regions of the parameter space (Sec.\,\ref{prospects:sigma}),
comparing the results with the sensitivity of dark matter detectors. 
We summarise our conclusions in
Section\,\ref{conclusions}.

\section{The low-energy NMSSM}
\label{NMSSM}

In this section we briefly review some of the most relevant aspects of
the NMSSM, and summarise the main constraints considered when
analysing the parameter space.

\subsection{Overview of the model}\label{overview}

With the addition of a gauge singlet superfield, the MSSM
superpotential is modified as follows: 
\begin{equation}\label{Wnmssm:def}
  W=
  \epsilon_{ij} \left(
  Y_u \, H_2^j\, Q^i \, u +
  Y_d \, H_1^i\, Q^j \, d +
  Y_e \, H_1^i\, L^j \, e \right)
  - \epsilon_{ij} \lambda \,S \,H_1^i H_2^j +\frac{1}{3} \kappa S^3\,,
\end{equation}
where $H_1^T=(H_1^0, H_1^-)$, $H_2^T=(H_2^+, H_2^0)$, $i,j$ are
$SU(2)$ indices, and $\epsilon_{12}=1$. 
In this model, the usual MSSM bilinear $\mu$ term is absent from the
superpotential, and only dimensionless trilinear couplings are present
in $W$. However, when the scalar component of $S$ acquires a VEV, an
effective interaction $\mu H_1 H_2$ is generated, with $\mu \equiv
\lambda \langle S \rangle$. Likewise, the soft SUSY breaking terms are
accordingly modified, so to include new soft-breaking masses for the
singlet ($m_{S}^2$), and additional trilinear couplings ($A_\lambda$
and $A_\kappa$). 
After electroweak (EW) symmetry is spontaneously broken, the neutral
Higgs scalars develop the following VEVs:  
\begin{equation}\label{vevs:DEF}
  \langle H_1^0 \rangle = v_1 \, , \quad
  \langle H_2^0 \rangle = v_2 \, , \quad
  \langle S \rangle = s \,. 
\end{equation}
After computing the tree-level scalar potential\footnote{For details
  about the Lagrangian and the neutral Higgs potential,
  $V^{\mathrm{Higgs}}_{\mathrm{neutral}}$, of the model see
  Ref.~\cite{Cerdeno:2004xw}.},
$V^{\mathrm{Higgs}}_{\mathrm{neutral}}$, one must ensure the presence
of a true minimum. The minimisation can be conveniently separated into
two steps. First, one imposes a minimum of
$V^{\mathrm{Higgs}}_{\mathrm{neutral}}$ with respect to the phases
(signs) of the VEVs, while in a later stage one derives the conditions
regarding the modulus of the VEVs. The first step immediately allows
to directly exclude combinations of signs for the parameters. In fact,
and working under the convention where $\lambda$ and $\tan \beta$ are
positive, and $\kappa$, $\mu(=\lambda\,s)$, $A_\lambda$ and $A_\kappa$
can have both signs, one can analytically show  that there are only
four distinct combinations of signs for $\kappa$, $A_\lambda$,
$A_\kappa$, and $\mu$ that ensure the presence of a minimum of
$V^{\mathrm{Higgs}}_{\mathrm{neutral}}$~\cite{Cerdeno:2004xw}:   
\begin{itemize}
\item[(i)] $\kappa>0$, $\mathrm{sign}(s)=\mathrm{sign}(A_\lambda)=
  -\mathrm{sign}(A_\kappa)$,
  
  which always leads to a minimum with respect to the phases of the
  VEVs. 
  
\item[(ii)] $\kappa>0$, $\mathrm{sign}(s)=-\mathrm{sign}(A_\lambda)=
  -\mathrm{sign}(A_\kappa)$,
  
  with $|A_\kappa| > 3 \lambda v_1 v_2 |A_\lambda|/(-|s
  A_\lambda| +\kappa |s^2|)$, where the denominator has to be
  positive.
  
\item[(iii)] $\kappa>0$, $\mathrm{sign}(s)=\mathrm{sign}(A_\lambda)=
  \mathrm{sign}(A_\kappa)$,
  
  with $|A_\kappa| < 3 \lambda v_1 v_2 |A_\lambda|/(|s
  A_\lambda| +\kappa |s^2|)$.
  
\item[(iv)] $\kappa<0$, $\mathrm{sign}(s)=\mathrm{sign}(A_\lambda)=
  \mathrm{sign}(A_\kappa)$,
  
  with  $|A_\kappa| > 3 \lambda v_1 v_2 |A_\lambda|/(|s  
  A_\lambda| -\kappa |s^2|)$, where the denominator has to be
  positive. 
\end{itemize}
The additional conditions regarding the minimisation of the potential
with respect to the Higgs VEVs can be derived, and these allow to
re-express the soft breaking Higgs masses in terms of $\lambda$,
$\kappa$, 
$A_\lambda$, $A_\kappa$, $v_1$, $v_2$ and $s$:
\begin{align}\label{2:minima}
  m_{H_1}^2 = & -\lambda^2 \left( s^2 + v^2\sin^2\beta \right)
  - \frac{1}{2} M_Z^2 \cos 2\beta
  +\lambda s \tan \beta \left(\kappa s +A_\lambda \right) \,,
  \nonumber \\
  m_{H_2}^2 = & -\lambda^2 \left( s^2 + v^2\cos^2\beta \right)
  +\frac{1}{2} M_Z^2 \cos 2\beta
  +\lambda s \cot \beta \left(\kappa s +A_\lambda \right) \,,
  \nonumber \\
  m_{S}^2 = & -\lambda^2 v^2 - 2\kappa^2 s^2 + \lambda \kappa v^2
  \sin 2\beta + \frac{\lambda A_\lambda v^2}{2s} \sin 2\beta -
  \kappa A_\kappa s\,,
\end{align}
where $v^2= v_1^2+v_2^2=2 M_W^2/g_2^2$ and $\tan \beta=v_2/v_1$.

The neutral Higgs spectrum in the NMSSM includes three scalars, and
two pseudoscalar states, whose mass matrices have been discussed
in~\cite{Cerdeno:2004xw}. For our present analysis, let us just recall
that in 
either case interaction and physical eigenstates can be related as 
\begin{equation}\label{2:Smatrix}
  h_a^0 \,=\, S_{ab} \,H^0_b\,\quad(a,b=1...3)\,,\quad \quad
  a^0_i \,=\, P_{ij} \,P^0_j\,\quad(i,j=1,2)\,,
\end{equation}
where $S$ ($P$) is the unitary matrix that diagonalises the $3 \times
3$ ($2 \times 2$) scalar (pseudoscalar) Higgs mass matrix. In both
sectors, we order the physical eigenstates as $m_{h_1^0} \lesssim
m_{h_2^0} \lesssim m_{h_3^0}$ and $m_{a_1^0} \lesssim m_{a_2^0}$. 
The singlet component of the lightest CP-even Higgs is therefore given
by $S_{13}$.

A final comment concerns the neutralino sector, now extended to
include the singlino component of the new chiral superfield, $S$. In
the weak interaction basis defined by ${\Psi^0}^T \equiv \left(\tilde
B^0=-i \lambda^\prime, \tilde W_3^0=-i \lambda_3, \tilde H_1^0, \tilde
H_2^0, \tilde S \right) $ the neutralino mass matrix 
reads  {\footnotesize \begin{equation}
    \mathcal{M}_{\tilde \chi^0} = \left(
    \begin{array}{ccccc}
      M_1 & 0 & -M_Z \sin \theta_W \cos \beta &
      M_Z \sin \theta_W \sin \beta & 0 \\
      0 & M_2 & M_Z \cos \theta_W \cos \beta &
      -M_Z \cos \theta_W \sin \beta & 0 \\
      -M_Z \sin \theta_W \cos \beta &
      M_Z \cos \theta_W \cos \beta &
      0 & -\lambda s & -\lambda v_2 \\
      M_Z \sin \theta_W \sin \beta &
      -M_Z \cos \theta_W \sin \beta &
      -\lambda s &0 & -\lambda v_1 \\
      0 & 0 & -\lambda v_2 & -\lambda v_1 & 2 \kappa s
    \end{array} \right).
    \label{neumatrix}
\end{equation}}
The above matrix can be diagonalised by means of a unitary matrix $N$,
\begin{equation}
  N^* \mathcal{M}_{\tilde \chi^0} N^{-1} = \operatorname{diag}
  (m_{\tilde \chi^0_1}, m_{\tilde \chi^0_2}, m_{\tilde \chi^0_3},
  m_{\tilde \chi^0_4}, m_{\tilde \chi^0_5})\,,
\end{equation}
where $m_{\tilde \chi^0_1}$ is the lightest neutralino mass. Under the
above assumptions, the lightest neutralino can be expressed as the
combination 
\begin{equation}
  \tilde \chi^0_1 = N_{11} \tilde B^0 + N_{12} \tilde W_3^0 +
  N_{13} \tilde H_1^0 + N_{14} \tilde H_2^0 + N_{15} \tilde S\,.
\end{equation}
In the following, neutralinos with $N^2_{13}+N^2_{14}>0.9$, or
$N^2_{15}>0.9$, will be referred to as Higgsino- or singlino-like,
respectively.

\subsection{Constraints on the parameter space}\label{constraints}

In addition to ensuring the presence of a minimum of the potential,
other constraints, both theoretical and experimental, must be imposed
on the parameter space generated by the low-energy NMSSM degrees of 
freedom 
\begin{equation}\label{low:energy:par}
\lambda, \, \kappa,\, \tan \beta,\, \mu,\, A_\lambda, \, A_\kappa\,.
\end{equation}
The soft supersymmetry-breaking terms, namely gaugino masses,
$M_{1,2,3}$, scalar masses, $m_{Q,L,U,D,E}$, and trilinear parameters,
$A_{Q,L,U,D,E}$, are also taken as free parameters and specified at
low scale.

A comprehensive analysis of the low-energy NMSSM phenomenology can be
obtained using the \nmh~2.0 code~\cite{Ellwanger:2005dv}. After
minimising the scalar potential, thus dismissing the presence of
tachyons and/or false minima, the Higgs boson masses are computed,
including 1- and 2-loop radiative corrections. Squark and slepton
masses are also calculated, as well as the corresponding 
mixing angles for the third generation. Chargino and neutralino masses
and mixings are evaluated and all the relevant couplings are derived.

Even though the general analysis is performed at low-energy, a further
theoretical constraint can be derived, 
namely the absence of a Landau pole 
for $\lambda$, $\kappa$, $Y_t$ and $Y_b$ below the GUT
scale. Including logarithmic one-loop corrections  
to $\lambda$ and $\kappa$, the latter constraint translates into
$\lambda \lsim 0.75$, $|\kappa| \lsim 0.65$, with $1.7 \lsim \tan
\beta \lsim 54$.

On the experimental side, \nmh~2.0 includes accelerator (LEP and
Tevatron) constraints, $B$-meson decays, and dark matter relic density
through a link to micrOMEGAS \cite{micro:old}.
In particular, direct bounds on the masses of the charged particles
($H^{\pm}$, $\tilde \chi^{\pm}$, $\tilde q$, $\tilde l$)  
and on the gluino mass are taken into account
\cite{LEPm,Tevm}. Excessive 
contributions to the invisible decay width of the $Z$ boson
\cite{pdg04,Z:inv}, as those potentially arising from $Z \to
\tilde\chi^0_i \tilde \chi^0_j $ and  
$Z \to h^0 a^0$, are also excluded from the parameter space.
Finally, in the neutral Higgs sector, one checks the constraints  on
the production rates for all the CP-even states $h^0$ and CP-odd
states $a^0$, in all the channels studied at LEP \cite{LEPH}: $e^+ e^-
\to h^0 Z$, independent of the $h^0$ decay mode (IHDM);  $e^+ e^- \to
h^0 Z$, dependent on the 
$h^0$ decay mode (DHDM), with the Higgs decaying via $h^0 \to b \bar
b$, $h^0 \to \tau^+ \tau^-$, 
$h^0 \to 2\, \text{jets}$ $h^0 \to \gamma \gamma$ and $h^0 \to
\text{invisible}$; associated production modes (APM), $e^+ e^- \to h^0
a^0$, with $h^0 a^0 \to 4 b$'s, $h^0 a^0 \to 4 \tau$'s and $h^0 a^0
\to a^0 a^0 a^0 \to 6 b$'s. 
In addition to the latter, one also  takes into account the possible
two body decays of all CP-even, CP-odd and charged Higgs bosons into
squarks and sleptons, as well as radiatively induced decays of neutral
Higgs bosons into two photons and two gluons.

Regarding $B$-meson decays, and although the \nmh~2.0 code already
contains a rough estimate of the $b\to s \gamma$ decay branching ratio
(evaluated at the leading order in QCD), we include in our code 
a more precise
computation of the $b\to s \gamma$ decay in the NMSSM~\cite{hiller},
taking into account next-to-leading order (NLO)
contributions~\cite{bsg_NLO,gambino-misiak}, following the results
of~\cite{kagan-neubert}. However, we only include leading order (LO) 
SUSY contributions to the Wilson coefficients at the $M_W$ 
scale\footnote{No charm-loop contributions were
  included in the analysis of  
  Ref.~\cite{kagan-neubert}, giving a SM central value of ${\rm BR}^{\rm
    SM}(b\to s \gamma)=3.293\times 10^{-4}$. 
  This result is obtained by extrapolating the value of the branching
  ratio evaluated at $\delta=0.9$ and  $\mu_b=m_b$, where $\delta$
  parameterises the photon energy  
  cut $E_{\gamma}> (1-\delta)m_b/2 $ and $\mu_b$ is the renormalisation
  scale. The corresponding new physics contribution has been implemented
  in our code by using the parameterisation of~\cite{kagan-neubert}
  evaluated 
  at $\mu_b=m_b$ and $\delta=0.9$.}.  
The calculation within the context of the MSSM at LO and NLO can be
found in
\cite{bbmr} and \cite{bsg_SUSY_NLO}, respectively. 
The most recent experimental world average for the branching ratio
(BR) reported by the Heavy Flavour Averaging Group
is~\cite{bsg_exp,pdg04}  
\begin{equation}\label{bsg:exp:value}
  {\rm BR}^{\rm exp}(b\to s \gamma)=(3.55 \pm 0.27)\times10^{-4}\,.
\end{equation}
On the other hand, the current SM prediction for the branching ratio
is~\cite{gambino}  
\begin{equation}\label{bsg:SM:value}
  {\rm BR}^{\rm SM}(b\to s \gamma)=(3.73 \pm 0.30)\times 10^{-4}\, ,
\end{equation}
where the charm-loop contribution has been
included~\cite{gambino-misiak}. 
We have estimated the theoretical error that results from varying the
scales in the $b\to s \gamma$ calculation
within the NMSSM, following the method described in
\cite{kagan-neubert}.  
We add to this the experimental error in quadrature. This procedure is 
performed at every point of the parameter space, typically leading to
an error of about $10\%$ of the total BR($b\to s \gamma$) value. 
Consistency at
$2\sigma$ 
with the experimental central value of
Eq.\,(\ref{bsg:exp:value}) is then demanded.

We have also included in our code other constraints coming from the
contribution of a light pseudoscalar $a^0$ in NMSSM to the rare $B$-
and $K$-meson decays~\cite{hiller}. When the pseudoscalar is very
light it could be produced in meson decays and significantly
affect the rates for $K-\bar{K}$ and $B-\bar{B}$ mixing and other SM
decays. 
In particular, our code takes into account the constraints from the
pseudo-scalar indirect contributions to  
$K-\bar{K}$ and $B-\bar{B}$ mixing,  $B\to \mu^+\mu^-$,  $B\to
X_s\mu^+\mu^-$, $B^-\to K^-\nu \bar{\nu}$, $B\to K_S^0 X^0$, and by the
direct production, at large $\tan{\beta}$, via $b\to s a^0$, $B\to K
a^0$, and $B\to \pi a^0$ decays.

Finally, in our analysis we will also include the
constraints coming from the SUSY contributions to the muon anomalous
magnetic moment, $a_{\mu}=(g_{\mu}-2)$~\cite{g-2_SUSY}. 
Taking into account the most recent theoretical predictions
for this quantity within the SM \cite{g-2_SM,newg2,kino} and the
measured experimental value \cite{g-2_exp}, the observed excess in
$a^{\rm exp}_{\mu}$ constrains a possible
supersymmetric contribution to be
$a_{\mu}^{\text{SUSY}}=(27.6\,\pm\,8)\times 10^{-10}$, where
theoretical and experimental errors have been 
combined in quadrature.

The evaluation of  $a_{\mu}^{\text{SUSY}}$ in the
NMSSM has been included in our analysis, and
consistency at the $2\sigma$ level imposed. Thus those regions of the
parameter space not fulfilling $11.6 \times 10^{-10} \lesssim
a_{\mu}^{\text{SUSY}} \lesssim 43.6 \times 10^{-10}$ will be
considered disfavoured.

\section{Dark matter in the NMSSM}
\label{DM:NMSSM}

The new features of the NMSSM have an impact on the 
properties of the lightest neutralino as a dark matter candidate,
affecting both its direct detection and relic abundance.

The computation of the spin-independent part of the 
neutralino-nucleon cross section was discussed
in detail in~\cite{Cerdeno:2004xw}. It was pointed out there that 
the existence of a fifth neutralino state, together with the presence
of new terms in the Higgs-neutralino-neutralino interaction
(which are proportional to $\lambda$ and $\kappa$), trigger new 
contributions to the spin-independent part of the
neutralino-nucleon cross section, $\sigma_{\tilde \chi^0_1 -p}$. 
On the one hand, although the
term associated with the $s$-channel squark exchange is formally
identical to the MSSM case, it can be significantly reduced if the
lightest neutralino has a major singlino composition. 
On the other hand, and more importantly, the dominant contribution to
$\sigma_{\tilde \chi^0_1 -p}$, associated to the exchange of CP-even
Higgs bosons on the $t$-channel can be largely enhanced when these are
very light.
In the NMSSM, the lightest CP-even Higgs can escape detection if its
singlet composition is large. For instance, this makes possible the
presence of scenarios with $m_{h_1^0}\lsim 70$~GeV, thus 
considerably
enhancing the neutralino-nucleon interaction.
Consequently, large detection cross sections can be obtained, 
even within the reach of the present
generation of dark matter detectors.

However, in order to be a good dark matter candidate, the lightest
NMSSM neutralino must also comply with the increasingly stringent
bounds on its relic density. Astrophysical constraints,
stemming from the analysis of 
galactic rotation curves~\cite{relic:curve}, clusters of galaxies and
large scale flows~\cite{relic:cluster}, suggest the following range
for the 
WIMP relic abundance 
\begin{equation}\label{om:mi}
  0.1 \lesssim \Omega h^2 \lesssim 0.3\,,
\end{equation}
which can be further reduced to 
\begin{equation}\label{om:wmap3}
  0.095 \lesssim \Omega h^2 \lesssim 0.112\,,
\end{equation}
taking into account the recent three years data from the WMAP
satellite~\cite{wmap}.

Compared to what occurs in the MSSM, one would expect several
alterations regarding the dominant processes. As discussed
in~\cite{Gunion:2005rw}, and as mentioned above regarding the direct
detection 
cross section, the differences can be present at distinct levels.
First, and given the presence of a fifth neutralino (singlino), the
composition of the annihilating WIMPs can be significantly different
from that of the MSSM in wide regions of the parameter space. Having
the possibility of a singlino-like lightest supersymmetric particle
(LSP), 
associated with the presence of new couplings in the interaction
Lagrangian, in turn favours  
the coupling of the WIMPs to a singlet-like Higgs, whose mass can be
substantially lighter that in the MSSM, given the more relaxed
experimental constraints. 
Regarding the channels through which neutralino annihilation occurs,
in the NMSSM we have new open channels,  
essentially due to the existence of light Higgs states. In summary,
the presence of additional Higgs states 
(scalar and pseudoscalar) favours annihilation via $s$-channel
resonances. On the other hand, having light $h^0_1$ and $a_1^0$ states
that are experimentally viable means that new channels with
annihilation  
into $Z\,h^0_1$, $h^0_1\,h^0_1$, $h^0_1\,a^0_1$ and $a^0_1\,a^0_1$
(either via $s$-channel $Z,\,h^0_i,\,a^0_i$ exchange or  
$t$-channel neutralino exchange) can provide important contributions
the annihilation and co-annihilation 
cross-sections~\cite{Belanger:2005kh}.

Noticing that important annihilation channels ($s$-channel) are
related to the $t$-channel processes responsible for the most relevant
contributions to $\sigma_{\tilde \chi^0_1 -p}$, one should expect a
strong interplay between a viable relic density, and promising
prospects for the direct detection of the NMSSM dark matter
candidate. 
In fact, there should be regions of the parameter space which provide
new and interesting scenarios\footnote{As concluded
  in~\cite{Gunion:2005rw}, it might even be possible to reconcile a
  very light neutralino with the experimental observations from
  DAMA, CDMS II, and WMAP.}.

\section{Results and discussion}
\label{discussion}

In this section, we study the viability of lightest NMSSM neutralino
as a good dark matter candidate. 
Motivated by the results obtained in~\cite{Cerdeno:2004xw}, we focus
on regions of the low-energy NMSSM parameter space where large direct
detection cross sections are likely to be obtained. 
Building upon the previous analysis, we apply the new constraints
(improved comparison with LEP and Tevatron data), $K$- and $B$-meson 
decays, $\asusy$, and compatibility with the WIMP relic density.
Finally, we discuss the prospects of the experimentally viable regions
regarding direct detection of dark matter.

Let us just recall that the free parameters of the model, associated
with the Higgs and neutralino sectors of the theory, 
are~\footnote{Although the soft gluino mass, $M_3$, is not directly
  related to the computation of the Higgs/neutralino masses and
  mixings, it plays a relevant role in contributing to the radiative
  corrections to the Higgs boson masses.} 
\begin{equation}
  \lambda\,, \quad \kappa\,,\quad \mu(=\lambda s)\,,\quad
  \tan \beta \,,\quad A_\lambda \,, \quad A_\kappa \,,\quad
  M_1\,, \quad M_2\,,\quad M_3\,.
\end{equation}
We assume that the gaugino mass parameters mimic, at low-energy, the
values of a hypothetical GUT unification ($\frac{M_3}{6}= M_1=
\frac{M_2}{2}$).

It should be
emphasised 
that in \nmh~2.0 some of the
input parameters are specified at a different scale than in the
former version \nmh~1.1~\cite{NMHDECAY}, which was used 
in the previous analysis \cite{Cerdeno:2004xw}. 
Although the difference
between the values of 
$\lambda$ (or $\kappa$) at the EW and the SUSY scales ($\approx$ 1
TeV) is very small, there is a substantial 
change in the value of the trilinear coupling $A_\lambda$. These
variations are induced by the top trilinear coupling $A_\text{top}$, 
and are approximately given by
$A_\lambda^\text{SUSY} \approx
A_\lambda^\text{EW}+A_\text{top}^\text{EW}$. 
Therefore, one needs to take this shift into account when comparing
the present results with those of \cite{Cerdeno:2004xw}.

Motivated by the results of~\cite{Cerdeno:2004xw} regarding the
prospects for direct detection of dark matter, we will be interested
in a regime of low $\tan \beta$, as well as in values of $|\mu|$ in
the range $110$~GeV~$\lesssim\,\mu\,\lesssim 200$~GeV (the lower limit
ensuring that in most cases one can safely avoid the LEP bound on
the lightest chargino mass). Likewise, the following intervals for
the trilinear couplings will
be taken: 
$-800$~GeV~$\lesssim\,A_\lambda\,\lesssim 800$~GeV, and
$-300$~GeV~$\lesssim\,A_\kappa\,\lesssim 300$~GeV (the
optimal ranges will typically correspond to 
$|A_\lambda|\,\sim 
400$~GeV and $|A_\kappa|\,\lesssim 200$~GeV, working in 
a small $\tan \beta$ regime).

Slepton and squark masses, as well as the corresponding trilinear
parameters, do not significantly affect the neutralino detection
properties, other than through the radiative corrections to the Higgs
masses. However, low-energy observables are very sensitive to their
specific values. In the following section we will see, for instance, 
how the experimental constraint on
$a_{\mu}^{\text{SUSY}}$ 
favours light sleptons.

As already done in Ref.~\cite{Cerdeno:2004xw}, we divide the scan of
the low-energy NMSSM parameter space following the results of the
minimisation with respect to the VEV phases, separately 
discussing each of the cases (i)-(iv) (see Section~\ref{overview}).

\subsection{NMSSM parameter space: updated constraints}
\label{constraints:up}

We first discuss the new constraints on the parameter space arising
from the improved analysis on the Higgs sector, 
the muon anomalous magnetic moment, and $K$- and
$B$-meson decays.

Among the new features implemented in \nmh~2.0, one finds additional
radiative corrections to the Higgs boson masses, including
corrections of order $g^2\, Y^2_{t,b}$ to the CP-even Higgs boson mass
(induced by stop/sbottom $D$-term couplings). 
Regarding the logarithmic one-loop corrections of the order $g^2$,
these are now dependent on the different  
masses of squarks/sleptons of distinct generations. Moreover, the
corrections to fourth order in $\lambda$ and $\kappa$ are also taken
into account. 
The computation of the sparticle spectrum is also complete in the new
version, and all squark and gluino data is confronted with the
constraints from both Tevatron and LEP. 
With respect to the results obtained in the previous
analysis~\cite{Cerdeno:2004xw}, the latter improvements only translate
into slight changes in the exclusion regions.

\begin{figure}[!t]
  \begin{center}
    \epsfig{file=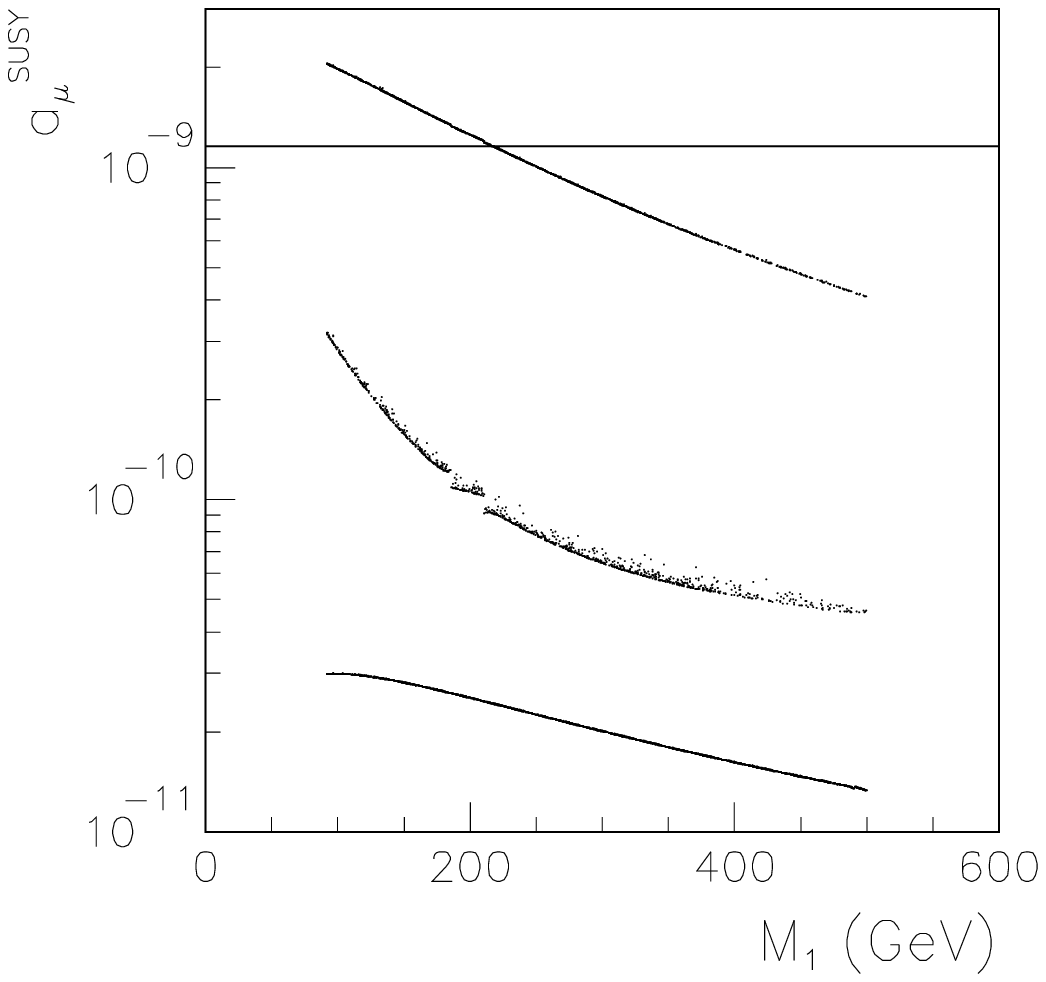, width=90mm}
  \end{center}
  \vspace*{-1cm}
  \captions{Supersymmetric contribution to the 
    anomalous magnetic moment of the muon as a
    function of the bino mass, 
    $M_1$, for $\tan \beta=5$, $\mu=150$ GeV, trilinear couplings
    in the range
    $-800 \lesssim A_\lambda \lesssim 800$ GeV, 
    $-300 \lesssim A_\kappa \lesssim 300$ GeV. 
    From bottom to top, the
    different bands correspond to the following values of the slepton
    mass and 
    lepton trilinear terms, 
    $m_{L,E}=1$~TeV with $A_E=1$~TeV,
    $m_{L,E}=150$~GeV with $A_E=1$ TeV, and 
    $m_{L,E}=150$~GeV with $A_E=-2.5$~TeV.
    A full scan on the $(\lambda,\kappa)$ plane
    has been performed for each case, including LEP and Tevatron
    experimental constraints. The horizontal solid line
    indicates the lower bound of the allowed
    $2\sigma$ interval.}
  \label{g-2} 
\end{figure}

Concerning the evaluation of the supersymmetric contributions to the
muon anomalous magnetic moment, the relevant processes comprise
neutralino-sneutrino as well as 
chargino-smuon loops. The only change
with respect to the MSSM is due to the fifth neutralino state and the
corresponding modified
neutralino-lepton-slepton coupling. Since we are interested in cases
with very low $\tan\beta$, the contributions from neutralino and
chargino loops are of similar magnitude, 
and very small. For example, with
$\tan\beta=3$ and slepton mass parameters above $m_{E,L}\sim1$ TeV one 
typically obtains $a_{\mu}^{\text{SUSY}}\sim10^{-11}$, which is
disfavoured. In order to obtain compatibility with the experimental
result an increase in the value of $\tan\beta$ is welcome, but this
would then lead to regions of the parameter space where, from the dark
matter point of view, the NMSSM resembles
the MSSM. The other possibility is decreasing the slepton (and
gaugino) masses. Furthermore, large values of the slepton trilinear 
couplings are needed 
in order to increase the $LR$ mixing in the smuon mass matrix. The
choice $\mu\,A_E<0$ is optimal, since it makes
the neutralino
contribution positive and large. 
For example, with $\tan\beta\sim 5$, $A_E=-2500$~GeV
and $m_{E,L}\lsim200$~GeV, one obtains $a_{\mu}^{\text{SUSY}} \gsim 
10^{-9}$ for $M_1\lsim215$~GeV. 
The relevance of these changes is illustrated in Fig.\,\ref{g-2},
where the numerical results for $\asusy$ are plotted versus the
bino mass, $M_1$, for different combinations of slepton mass and trilinear
couplings. For each case we have also varied $A_\lambda$ and
$A_\kappa$ over a wide range and scanned the whole
$(\lambda,\kappa)$ plane, which as evidenced in the figure has
virtually no effect on the resulting $\asusy$.
We have also included the various LEP and Tevatron constraints. 
For the rest of our analysis, we will assume
$m_{E,L}=150$~GeV and $A_E=-2500$~GeV. Also, and 
unless otherwise stated, we will set the bino mass to $M_1=160$ GeV,
which, according to 
Fig.\,\ref{g-2}, leads to a sufficiently large
$\asusy$. 
The detection properties of the neutralino are in general
quite insensitive to changes in the slepton sector. 
Notice however that if one does not wish to impose
the bound on the muon anomalous magnetic moment, 
heavy sleptons (equal to squarks) 
can be taken which would not
affect the dark matter predictions.

Regarding the bounds arising from $K$- and $B$-meson physics, the most
important role is played by the $b \to s\,\gamma$ decay, which can in
principle exclude important regions of the parameter space.
Concerning the other $K$- and $B$-meson processes discussed in
Section~\ref{constraints}, 
we have verified that throughout the investigated NMSSM parameter
space they are always in good agreement with experiment, so that we
will make no further reference to the latter in the following
discussion of the numerical results.

In the present analysis we will not take into account any source of
flavour violation other than the 
Cabibbo-Kobayashi-Maskawa (CKM) matrix. Moreover, we will be
systematically considering large values for the gluino mass (above
1~TeV). Under the latter assumptions, the most important contributions
to BR($b \to s\,\gamma$) arise in general from charged Higgs and
chargino mediated diagrams \cite{bbmr}.

On the one hand,
when the dominant contributions are
those stemming from charged Higgs exchange, the results for BR($b \to
s\,\gamma$) closely follow the behaviour of the charged Higgs mass,
which in the NMSSM is given by 
\begin{equation}\label{charg:higgs:mass}
  m_{H^\pm}^2\;=\;\frac{2 \,\mu^2}{\sin(2 \beta)}\,
  \frac{\kappa}{\lambda}\,-\, v^2
  \,\lambda^2 \,+\, \frac{2 \,\mu \,A_\lambda}{\sin(2 \beta)}
  \,+\, M_{W}^2\,.
\end{equation}
From the above, we expect that smaller values of BR($b \to s\,\gamma$)
should be obtained for large $m_{H^\pm}^2$, and therefore when
$\kappa/\lambda$ is sizable (for positive values of
$\kappa$) or for small $\kappa/\lambda$ (if $\kappa < 0$). In general, 
smaller values of the BR($b \to s\,\gamma$) will be also associated to
larger values of the product $\mu \,A_\lambda$. Furthermore, the
leading term of the 
Wilson coefficient associated to the charged Higgs varies as
$\tan^{-2}\beta$ \cite{bbmr}. 
As a consequence, one expects a decrease of this
contribution as $\tan\beta$ increases.  
On the other hand, in a regime of $\mu \lesssim M_2$, the lightest
chargino is Higgsino-dominated, so that its mass is
also quite small ($m_{\tilde\chi^\pm_1} \sim \mu$). Thus, the chargino
contributions (which are opposite in sign to those of the charged
Higgs) are also expected to play a relevant role, 
although, in the cases analysed in this paper ($\tan\beta\lsim
10$), they are not dominant. 
Gluino contributions are also very small, given the little 
flavour 
violation in the down squark sector, and the sizable values of
$M_3$. Likewise, neutralino exchange contributions are almost
negligible. 
We thus find that, in general, the NMSSM contribution to BR($b \to
s\,\gamma$) at low $\tan\beta$ 
is large and mostly arising from charged Higgs loops. This leads to
stringent constraints on the parameter space.

Let us study the effect of the experimental bound on BR($b \to
s\,\gamma$), together with the updated accelerator constraints on the
NMSSM parameter space.
After this 
first survey
we will no longer separately address 
the $K$- and $B$-meson constraints (and $a_\mu$)
from those arising from LEP/Tevatron data. Henceforth, experimentally
allowed regions will be those that not only comply with the latter
data, but that also exhibit BR($b \to s\,\gamma$) within $2\sigma$
from its central experimental value.
As mentioned before, 
in order to satisfy the constraint on the muon anomalous magnetic
moment, we take $M_1=160$ GeV in the following subsection, for which,
in the case with  $\tan\beta=5$, 
$\asusy\approx1.4\times10^{-9}$ (see
Fig.\,\ref{g-2}).

\subsubsection{$\mu A_{\kappa}<0$  and $\mu A_{\lambda}>0$
  ($\kappa>0$)}\label{mp}

As discussed in~\cite{Cerdeno:2004xw}, this is one of the most
interesting areas of the parameter space, since 
although sizable regions are 
excluded due to the occurrence of tachyons in the CP-even
Higgs sector (namely for larger values of $|A_\kappa|$), the
possibility of having experimentally viable light scalar Higgs leads
to potentially large values for $\sigma_{\tilde \chi^0_1 -p}$.

\begin{figure}[!t]
  \hspace*{-7mm}
  \epsfig{file=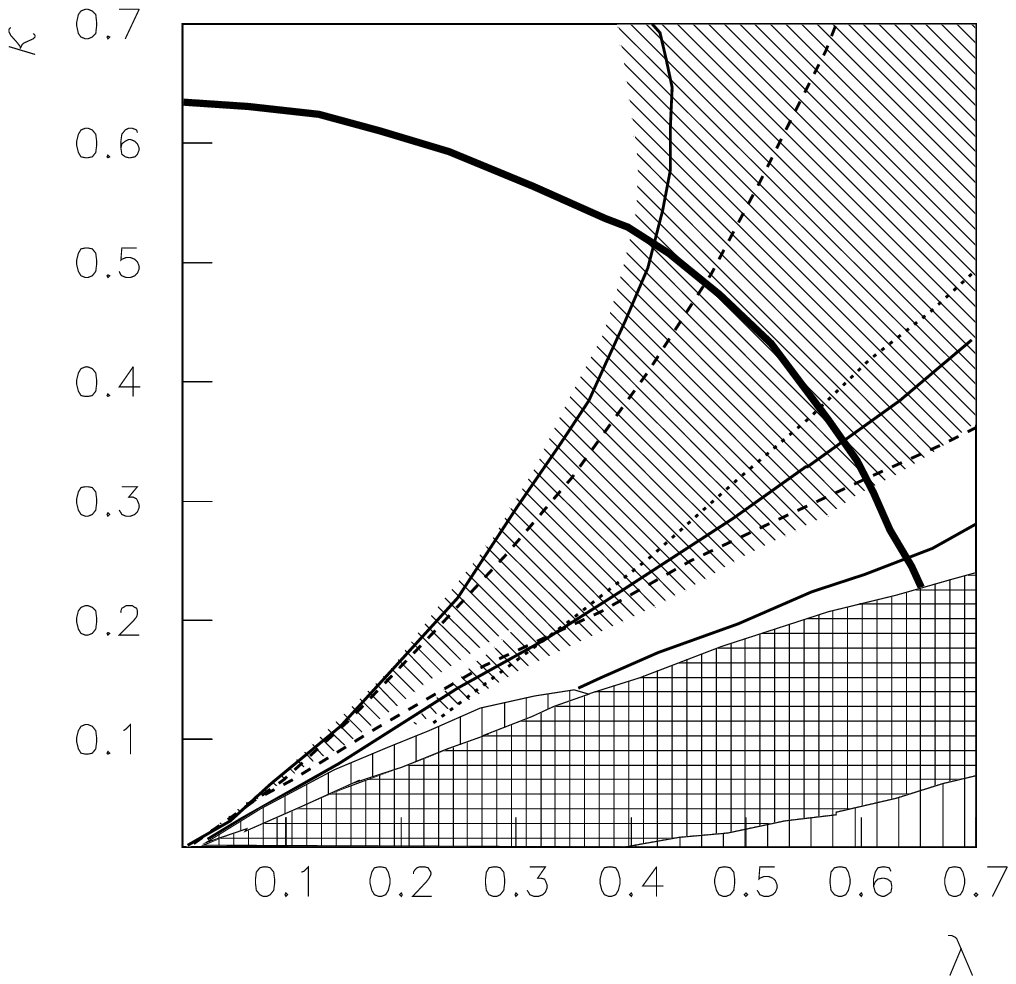, 
    width=90mm}\hspace*{-10mm}
  \epsfig{file=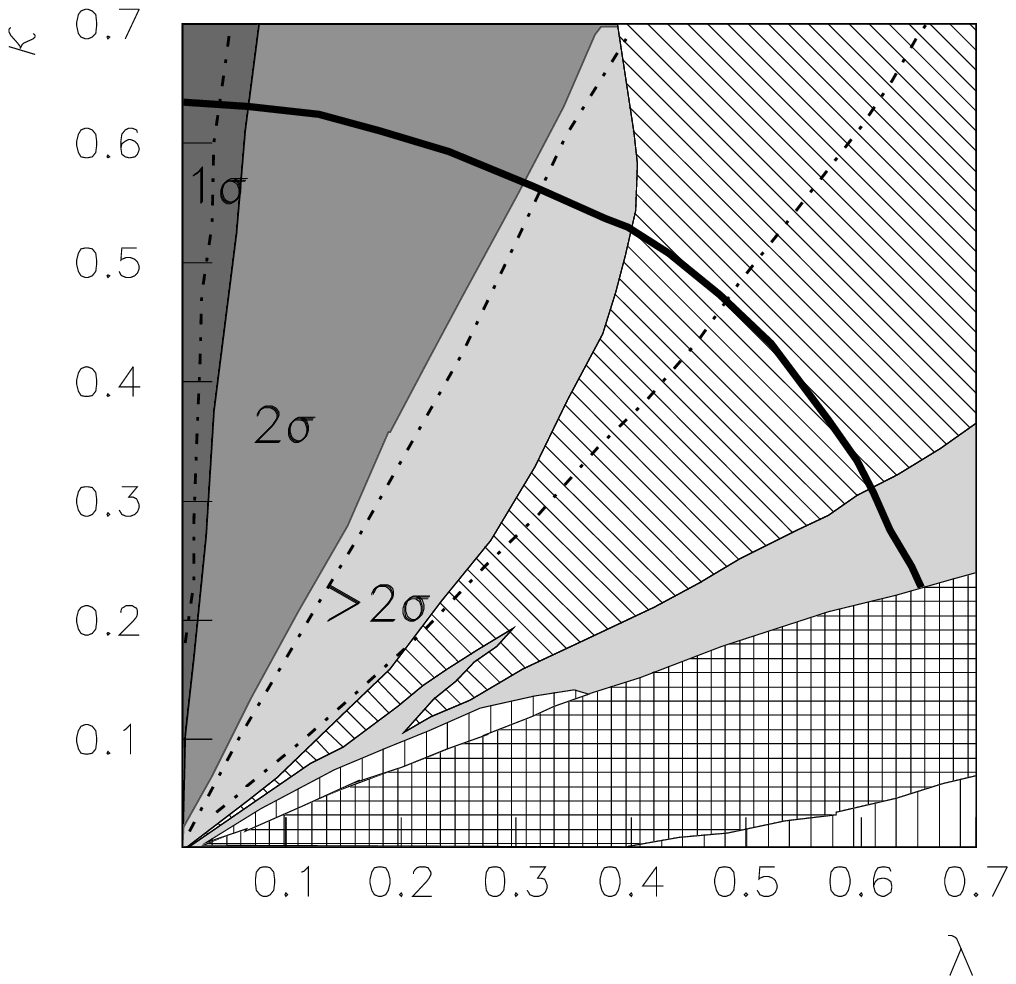,
    width=90mm}  
  \vspace*{-1cm}
  \captions{Effect of the experimental constraints on the
    $(\lambda,\kappa)$ plane for an example with $\tan \beta =3$,  
    $A_{\lambda}=200$~GeV, $A_{\kappa}=-200$~GeV and $\mu=130$~GeV. In 
    both cases, the gridded area is excluded due to the appearance of
    tachyons, while the vertically ruled area corresponds to the
    occurrence of unphysical minima. The oblique ruled area is
    associated with points that do not satisfy the LEP and/or
    Tevatron constraints or where (at least) the LEP bound on direct
    neutralino production is   violated. The region above the thick
    black line is disfavoured due to the occurrence of a Landau
    pole below the GUT scale. On the 
    left plot, from top to bottom, solid lines indicate different
    values 
    of the lightest Higgs mass, $m_{h_1^0}=114,\,75,\,25$~GeV.
    Dashed lines separate the regions where the lightest scalar 
    Higgs has a singlet composition given by
    ${S_{13}^{\,2}}=0.1,\,0.9$ (from top to bottom). 
    Finally in the area below the dotted line, the lightest
    neutralino has a singlino composition greater than
    $N_{15}^2=0.1$. On the right, grey areas represent the
    theoretical predictions for BR($b \to s\,\gamma$). From left to
    right, $1\sigma$ (dark), $2\sigma$ (medium) and excluded
    (light) regions are shown. Dot-dashed lines stand for the
    different values of the charged Higgs mass, 
    $m_{H^\pm}=1000,\, 500, \, 450$~GeV (from left to right).} 
  \label{fig:kl_160_200_130_-200_3}
\end{figure}

As an example, we represent on the left-hand side of 
Fig.\,\ref{fig:kl_160_200_130_-200_3} the
$(\lambda,\kappa)$ parameter space for an example with $\tan \beta
=3$, $A_{\lambda}= 200$~GeV, $A_{\kappa}=-200$~GeV and
$\mu=130$~GeV. The tachyonic region in the CP-even Higgs sector is
depicted, as well as the region excluded due to the 
presence of false minima of the
potential. 
An important part of the
theoretically allowed region is also ruled out due to conflict with
LEP and/or Tevatron data. 
This owes to the fact that the doublet component of the lightest
scalar Higgs is very large and gives rise to excessive Higgs
production rates, 
in particular, $e^+e^- \to h^0 Z$, IDHM and DHDM ($h^0 \to b \bar b$
and to a lesser extent, $h^0 \to 2$~jets).  
Once all these bounds are applied, 
a small area on the right of that experimentally excluded
survives, remarkably exhibiting very light Higgses and neutralinos
(associated with a singlet/singlino component above 90\%) and therefore
clearly characteristic of the NMSSM. 
We recall that these are the regions where one expects to find large 
theoretical predictions for $\crosssec$.

On the right-hand side of Fig.\,\ref{fig:kl_160_200_130_-200_3} we
superimpose the results for the BR($b \to s\,\gamma$) on the 
$(\lambda,\kappa)$ plane. 
As discussed in the previous section, the resulting branching ratio
is typically large, BR($b \to
s\,\gamma$)$\gsim3.5\times10^{-4}$, and increases to as much as
$\sim5\times10^{-4}$
in regions with
small $\kappa/\lambda$, where the charged Higgs mass is smaller.
Notice therefore that the regions of the
$(\lambda,\kappa)$ plane associated with larger values of BR($b \to 
s\,\gamma$) typically correspond to those where the
largest predictions for $\sigma_{\tilde \chi^0_1 -p}$ are found. 
In this example, only a 
small triangular region with $\lambda\lsim0.05$, for $\kappa<0.7$, is
within a $1\sigma$ deviation from the experimental bound of
Eq.\,(\ref{bsg:exp:value}) and $\lambda\lsim0.35$ is needed in order to
be 
within $2\sigma$ of that result.  
In the plot we also indicate with dot-dashed lines the different
values of the charged Higgs mass, thus illustrating the
correlation between its decrease and the increase in BR($b \to
s\,\gamma$).

\begin{figure}[!t]
  \hspace*{-7mm}
  \epsfig{file=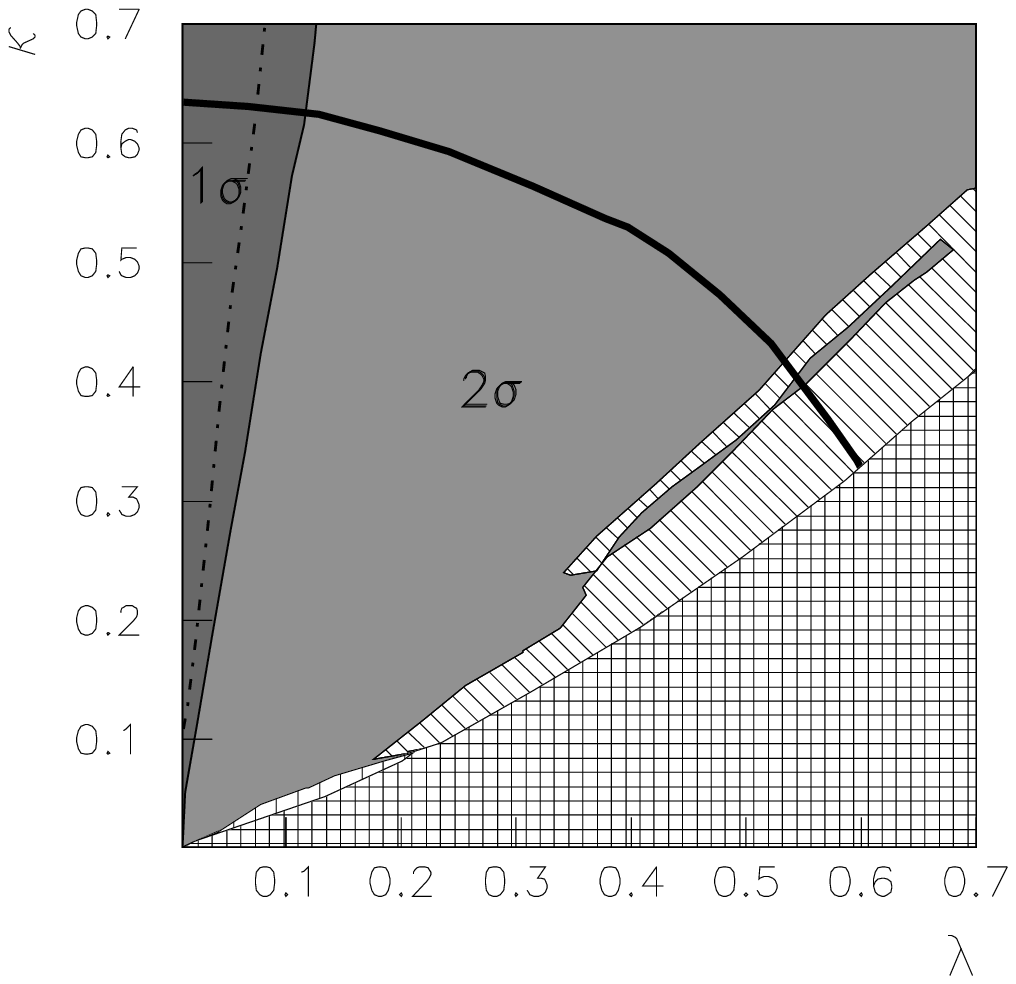,
    width=90mm}\hspace*{-10mm}
  \epsfig{file=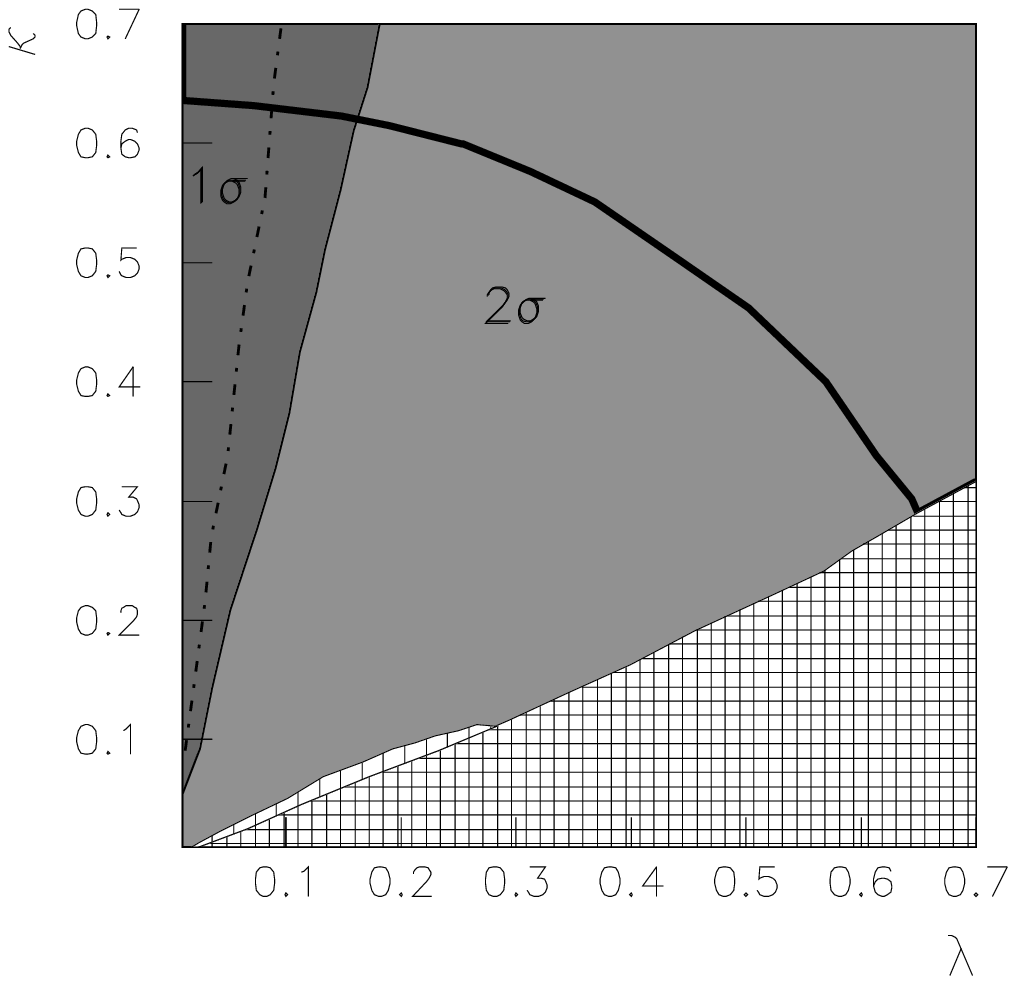,
    width=90mm}  
  \vspace*{-1cm}
  \captions{$(\lambda,\kappa)$ parameter space for $\tan \beta =5$, 
    $A_{\kappa}=-200$~GeV, and $\mu=130$~GeV. On the left we take
    $A_{\lambda}=200$~GeV, while on the right we consider
    $A_{\lambda}=400$~GeV. Line and colour code follow the
    conventions of  
    Fig.\,\ref{fig:kl_160_200_130_-200_3}. In this case, the single
    dot-dashed line corresponds to $m_{H^\pm}=1000$ GeV.}
  \label{fig:kl_160_200y400_150_-200_5} 
\end{figure}

The effect of the various experimental constraints on the parameter
space is very sensitive to variations in the input parameters. We will
now investigate how changes in $\tan\beta$ and $A_\lambda$ affect the
resulting BR($b \to s\,\gamma$). 
On the one hand, as already mentioned, 
increasing the value of $\tan\beta$ leads to a
reduction of the charged Higgs contribution. Since in our case 
this is the leading contribution to BR($b \to s\,\gamma$), an
enhancement in $\tan\beta$ enlarges the regions of the parameter space
which are consistent with the experimental constraint. 
This is illustrated on the left-hand side of 
Fig.\,\ref{fig:kl_160_200y400_150_-200_5} with the same example of
Fig.\,\ref{fig:kl_160_200_130_-200_3}, but now taking
$\tan\beta=5$. The resulting charged Higgses are heavier
($m_{H^\pm}>500$ GeV) and as a consequence the 
entire $(\lambda,\kappa)$ plane fufils 
the experimental constraint on BR($b \to s\,\gamma$). Notice that LEP
and Tevatron constraints are also modified.
On the other hand, an increase in the trilinear term $A_{\lambda}$
also leads to heavier charged Higgses as seen in
Eq.\,(\ref{charg:higgs:mass}). Therefore, this 
can induce a further decrease in BR($b\to 
s\,\gamma$). An example of this is
shown on the right-hand side of
Fig.\,\ref{fig:kl_160_200y400_150_-200_5}, where in addition to
$\tan\beta=5$, $A_{\lambda}=400$~GeV has been used. Again, the
whole $(\lambda,\kappa)$ plane is allowed due to the increase in
$m_{H^\pm}$.

\begin{figure}
  \begin{center}
    \epsfig{file=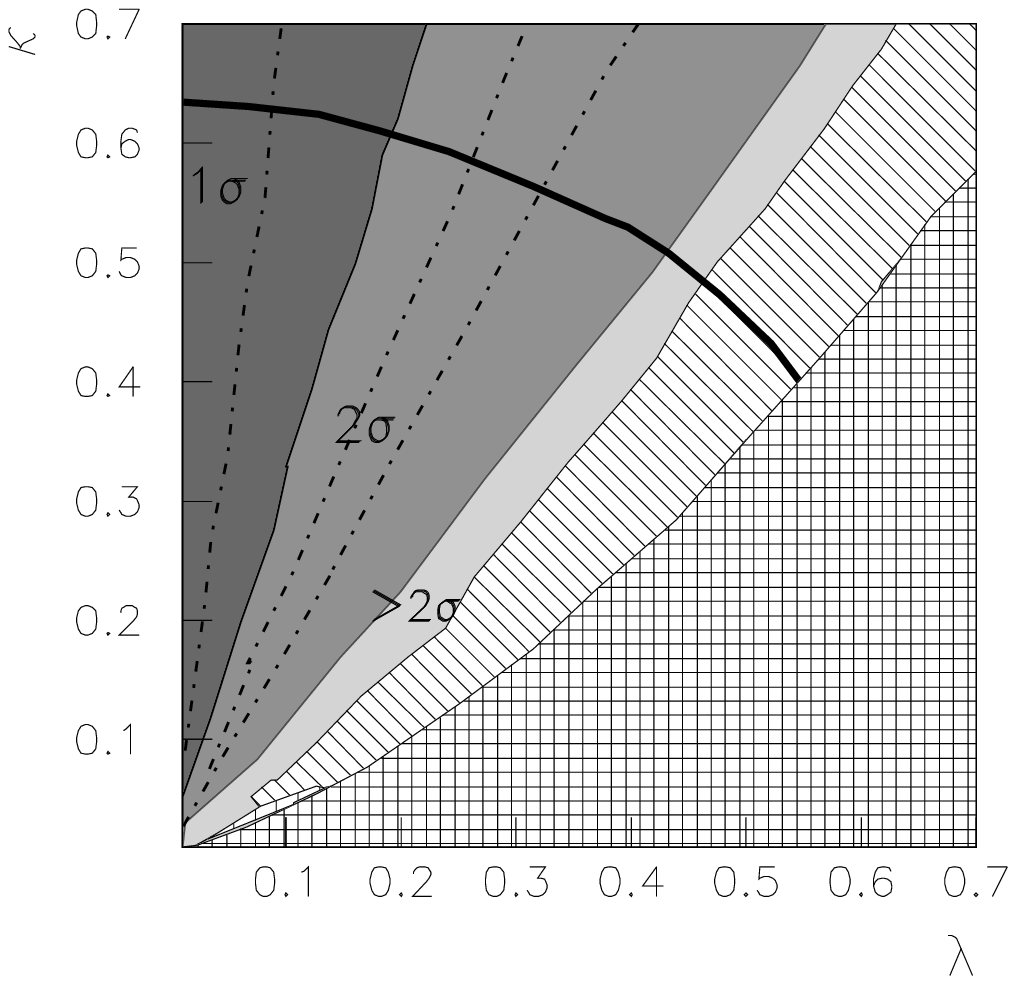,
      width=90mm}  
  \end{center}
  \vspace*{-1cm}
  \captions{$(\lambda,\kappa)$ parameter space for $\tan
    \beta =5$,   $A_{\lambda}=-200$~GeV, $A_{\kappa}=200$~GeV and
    $\mu=-130$~GeV. 
    Line and colour code following the conventions of
    Fig.\,\ref{fig:kl_160_200_130_-200_3}.} 
  \label{fig:kl_160_-200}
\end{figure}

Let us finally comment on the possibility of changing the signs of
$\mu$, $A_\lambda$, and $A_\kappa$, while keeping  $\mu A_\lambda >0$
and $\mu A_\kappa<0$.
Although the Higgs
potential is invariant under this change, the same
does not occur for the chargino and neutralino sectors, so that both
these spectra, as well as the experimental constraints are likely to
be modified. As an illustrative example, we present in
Fig.\,\ref{fig:kl_160_-200} the same case as in the left-hand side of 
Fig.\,\ref{fig:kl_160_200y400_150_-200_5} but with the
opposite signs for $\mu$, $A_\lambda$, and $A_\kappa$. 
There are some important alterations to the areas excluded by 
unphysical minima and
experimental constraints, both of which are now more extensive.
Finally, notice that the BR($b \to s\,\gamma$) now excludes a larger
area of the $(\lambda,\kappa)$ plane, thereby disfavouring those areas
which potentially lead to larger neutralino detection cross sections.

In the light of this analysis, the optimal areas of the parameter
space correspond to those with $\mu,\,A_\lambda>0$, and $A_\kappa<0$, 
and where $\tan\beta$ and $A_\lambda$ are relatively large.
In
order to keep within the context where NMSSM-like dark matter
scenarios can be obtained, we will use
$\tan\beta\le5$.

\subsubsection{$\mu A_{\kappa}<0$ and $\mu A_{\lambda}<0$
  ($\kappa>0$)} 
\label{mm}

\begin{figure}
  \begin{center}
  \epsfig{file=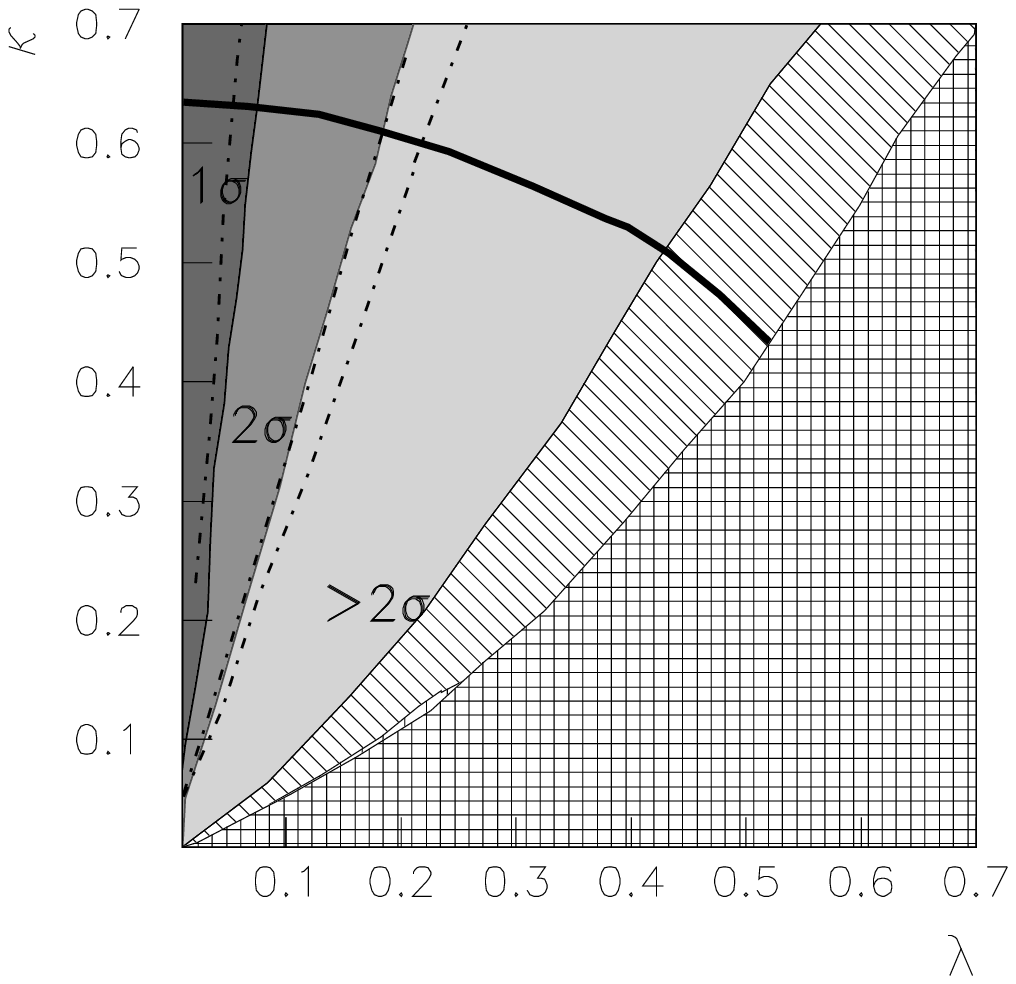,
    width=90mm}  
  \end{center}
  \vspace*{-1cm}
  \captions{
    $(\lambda,\kappa)$ parameter space for $\tan \beta
    =5$,   $A_{\lambda}=-200$~GeV, $A_{\kappa}=-200$~GeV and
    $\mu=130$~GeV. 
    Line and colour code following the conventions of
    Fig.\,\ref{fig:kl_160_200_130_-200_3}.} 
  \label{fig:kl_160_-200b}
\end{figure}

Compared to the previous case, the presence of tachyons gives rise to
far stronger constraints. Occurring now in both CP-even and CP-odd
Higgs sectors, the non-physical (tachyonic) 
solutions exclude very large areas of
the parameter space. 
Regarding the LEP experimental exclusions,
these arise from excessive contributions to $h^0 \to b \bar b$ and
$h^0 \to 2$ jets, and cover an area wider than what had been previously 
identified in~\cite{Cerdeno:2004xw} (a consequence of the improved
computation of the Higgs spectrum). 
In addition, due to the lightness of the charged Higgs bosons, 
an important region is also excluded due to very large 
BR($b \to s \,\gamma$), so that the only surviving regions are those
associated with $\lambda \lesssim 0.2$. 
As an example, 
Fig.\,\ref{fig:kl_160_-200b} displays a case with 
$\tan \beta=5$,   $A_{\lambda}=-200$~GeV, $A_{\kappa}=-200$~GeV and
$\mu=130$~GeV.  
Let us remark that since the lower-right
corner of the $(\lambda,\kappa)$ plane is not accessible, one cannot
find light neutral Higgs states, so that interesting 
prospects regarding the direct detection of dark matter should not be expected.

Although varying the several parameters results in modifications of
the excluded areas (LEP/Tevatron, $b \to
s\,\gamma$ and unphysical minima), in all cases these are sizable. 
Only very reduced
regions, corresponding to small values of $\lambda$
survive all the constraints. 
In these areas the singlet component of the lightest Higgs is
negligible and the lightest neutralino is Higgsino-like, therefore
resembling MSSM scenarios. 
The complementary region, with $A_{\lambda},\,A_{\kappa}>0$ and
$\mu<0$, leads to even more extensive tachyonic regions, and we will
not further discuss it.

\subsubsection{$\mu A_{\kappa}>0$ and $\mu A_{\lambda}>0$
  ($\kappa>0$)} 
\label{pp}

This combination of signs leads
to a parameter space which is plagued with tachyons
\cite{Cerdeno:2004xw}, arising from both
Higgs sectors. Contrary to what was noticed for the previous cases,
here the unphysical minima occur for small values of $\lambda$.
The remaining areas in the 
$(\lambda,\kappa)$ plane are also very affected by experimental
constraints. 
As a consequence, only very reduced areas of the parameter space
survive.

\begin{figure}
  \hspace*{-7mm}
  \epsfig{file=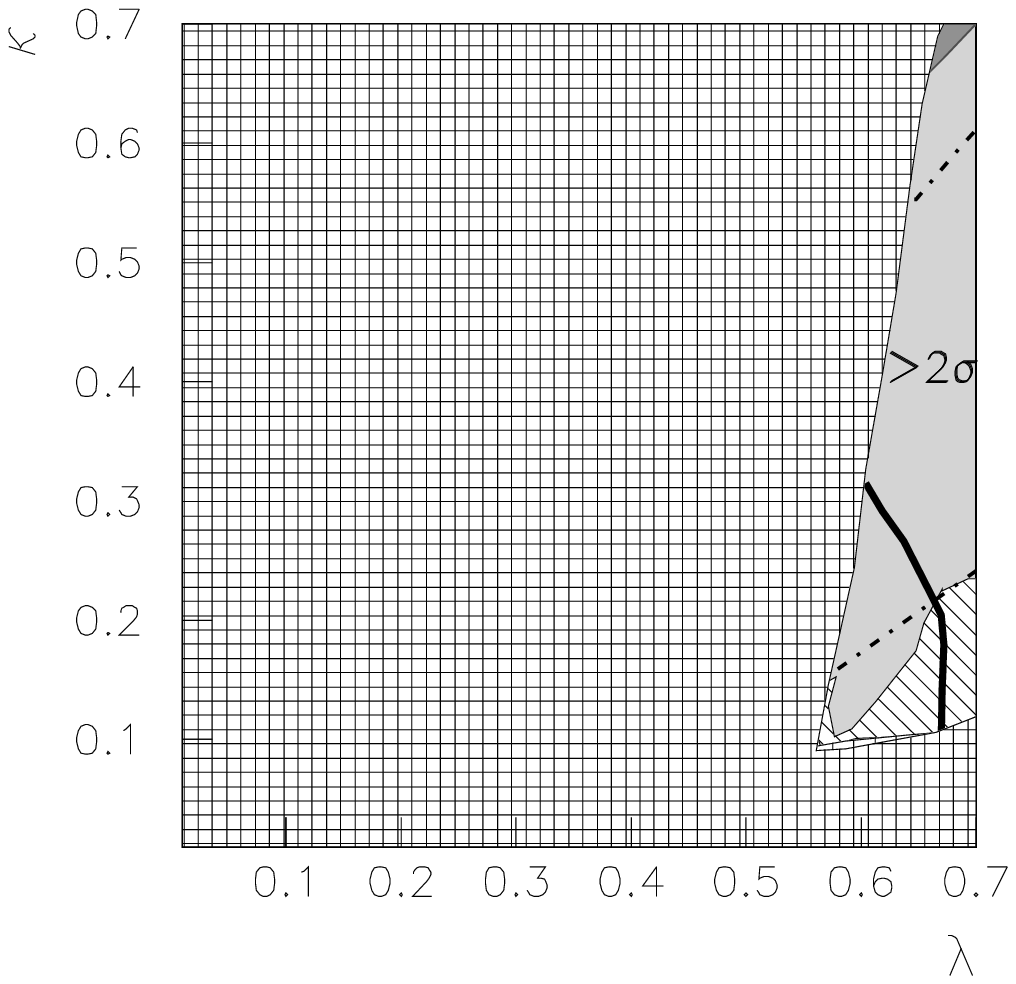, 
    width=90mm}\hspace*{-10mm}
  \epsfig{file=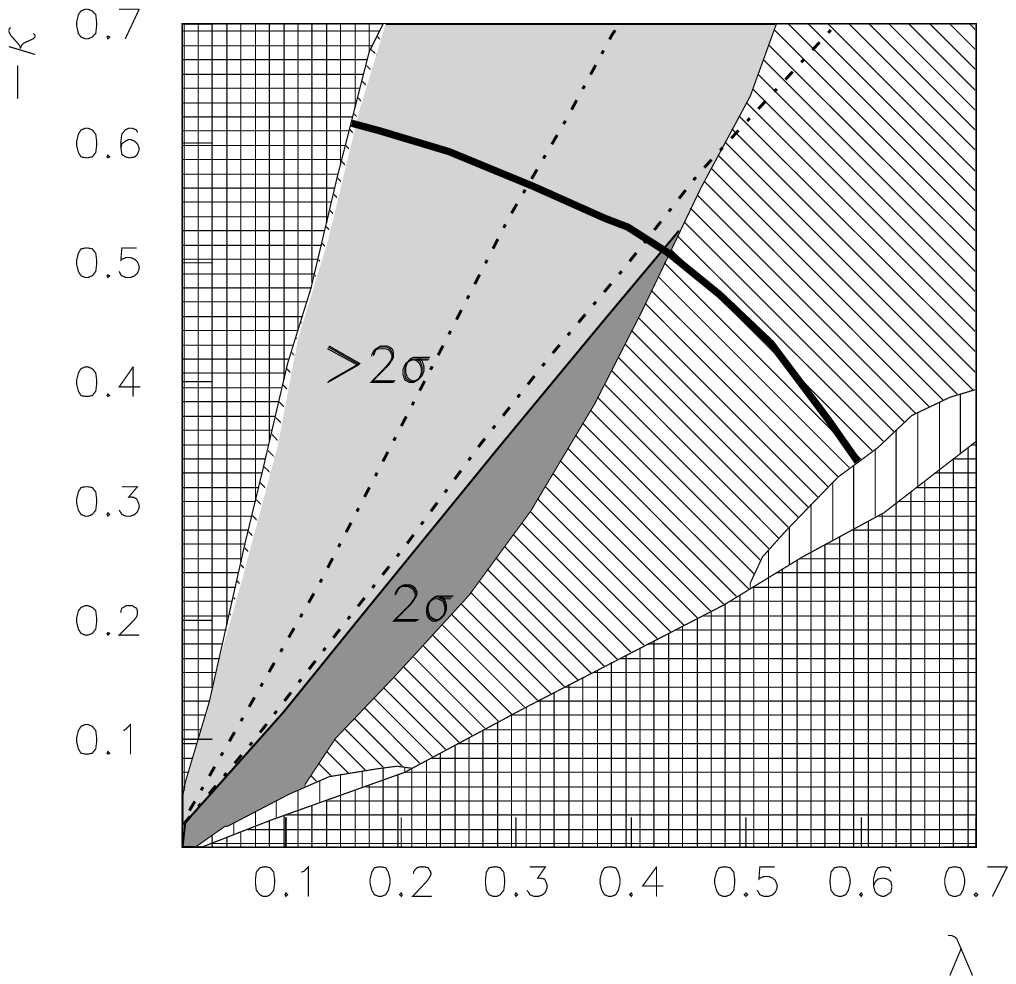, 
    width=90mm} 
  \vspace*{-1cm}
  \captions{On the left $(\lambda,\kappa)$ parameter space for $\tan
    \beta =3$, 
    $A_{\lambda}= 200$~GeV, $A_{\kappa}=50$~GeV and $\mu=160$~GeV. 
    On the right $\tan \beta =5$, $A_{\lambda}= 400$~GeV,
    $A_{\kappa}=200$~GeV, and  
    $\mu=130$~GeV for $\kappa<0$. 
    Line and colour code following the conventions of
    Fig.\,\ref{fig:kl_160_200_130_-200_3}.}\label{fig:kl+} 
\end{figure}

For example, considering the choice $A_{\lambda}= 200$~GeV,
$A_{\kappa}=50$~GeV and $\mu=160$~GeV, with $\tan \beta=3$, we observe
that once the areas corresponding to the occurrence of tachyons are
excluded, the small surviving region is still plagued by false minima
as well as by the violation of several experimental constraints. In
particular, lower values of $\kappa$ are ruled out due to conflict
with LEP ($h^0 \to b \bar b$ and  
$h^0 \to 2$ jets) and excessive contributions to the BR($b \to
s\,\gamma$).  
This is illustrated on the left-hand side of Fig.\,\ref{fig:kl+}.

Although reducing the values of $A_{\kappa}$ and $\tan\beta$ 
enlarges the areas where
physical minima can be found, the addition of the experimental
constraint on BR($b \to s \gamma$) to the LEP and Tevatron bounds
typically rules out the whole $(\lambda,\kappa)$ plane. 
Increasing $\tan\beta$ in order to reduce the contribution to BR($b
\to s \gamma$) is also highly disfavoured since the tachyonic region
becomes more important.
As a consequence, no interesting implications for neutralino 
dark matter detection are expected in this case.

\subsubsection{$\mu A_{\kappa}>0$ and $\mu A_{\lambda}>0$
  ($\kappa<0$)} 
\label{nkpp}

The only viable combination of
signs associated with negative values of $\kappa$ is also plagued by
the appearance of tachyons, both in the CP-even
and CP-odd Higgs
sectors, towards the regions with small $\lambda$
\cite{Cerdeno:2004xw}.  
In addition, experimental constraints also exclude large portions of
the parameter space in the vicinity of the tachyonic regions. 
As an example, let us mention that for the case $A_{\lambda}=
100$~GeV, $A_{\kappa}=50$~GeV and $\mu=130$~GeV, with 
$\tan \beta=3$, all the parameter space associated with physical
minima is ruled out, since either DHDM constraints ($h^0 \to b
\bar b$ and $h^0 \to 2$ jets) are violated or 
consistency with the BR($b \to s\,\gamma$) bound is not achieved. 
Increasing $A_\lambda$, $A_\kappa$, and $\tan \beta$ leads to a
significant improvement. 
For example, with $A_{\lambda}= 400$~GeV, $A_{\kappa}=200$~GeV,
$\mu=130$~GeV with $\tan \beta=5$ some allowed areas are found, as
shown on the right-hand side of Fig.\,\ref{fig:kl+}. 
Nevertheless, 
LEP/Tevatron experimental constraints together with the bound on
BR($b \to s\,\gamma$) rule out those parts of the parameter space
where the Higgs is light and more singlet-like, and in which
$\crosssec$ can be sizable. 
The remaining 
allowed regions correspond to a rather small 
area, in which the lightest Higgs is essentially doublet-like, while
the lightest neutralino exhibits a strong Higgsino  
dominance.

\subsection{Neutralino relic density}
\label{constraints:om}

The next step in our analysis is to take into account the available
experimental data on the WIMP relic density. In order to be a viable
dark matter 
candidate, the lightest NMSSM neutralino must have an abundance
within the ranges presented in Eqs.~(\ref{om:mi},\ref{om:wmap3}).
Similar to what occurs in the MSSM, this additional
constraint further reduces the regions of the low-energy
parameter space. Moreover, and as hinted before, one expects that
$\relic$ will in general lie below the experimental ranges.

A thorough analysis of the relic density of dark matter in the NMSSM
has been carried out in~\cite{Belanger:2005kh}. It was found
that compatibility with the WMAP constraint is possible in
the regions where the lightest Higgs is dominated by the
doublet components and the lightest neutralino is a bino-Higgsino
mixture. Apart from 
possible Higgs resonances, compatible values of $\relic$ are found
for two distinct regions: $\mu \gg M_2$ and $\mu \gtrsim M_1$ (similar
to the MSSM for small to intermediate values of $\tan \beta$). Also 
note that, for regions with $\mu \lesssim M_1$,
Higgsino-singlino neutralinos with masses below $M_W$ can
give a relic density within the WMAP range (or larger) essentially
because the annihilation into $Z$ and $W$ gauge bosons is kinematically
forbidden. 
It is also worth noticing that a pure bino LSP also offers interesting
scenarios, with a remarkable role being played by $s$-channel Higgs
resonances (else $\relic$ tends to be
above the experimental bound). Additional LSP annihilation via
scalar or pseudoscalar Higgses can also play a relevant role.

In order to understand the results for $\relic$ one needs to take into
account the variations in the mass and composition of the lightest
neutralino in the $(\lambda,\kappa)$ plane, as well as in the Higgs
sector. In general, the neutralino relic density will be too
small in those regions of the parameter space where it is
Higgsino-like and increases when the neutralino becomes more
singlino-like. In addition, one should consider
the possible existence of
resonant annihilation 
(when twice 
the neutralino mass equals the mass of
one of the mediating particles in an $s$-channel) 
and the kinematic thresholds for the
various channels (e.g., annihilations into $ZZ$, $WW$, $Zh_i^0$,
$h_i^0h_j^0$, $a_i^0a_j^0$, and $a_i^0h_j^0$).

Since the goal of our present study is to
discuss the potential of 
NMSSM-like scenarios regarding the theoretical predictions for
$\crosssec$, 
in this subsection  
we focus on those 
examples of the parameter space having
large neutralino detection cross section.
We investigate to which extent the inclusion of
the bound of the relic density further constrains the parameter space. 
As pointed out in~\cite{Cerdeno:2004xw}, these scenarios typically
occur in association to singlet-like $h^0_1$, with
singlino-Higgsino neutralinos. Let us study one example in detail.

\begin{figure}
  \hspace*{-7mm}
  \epsfig{file=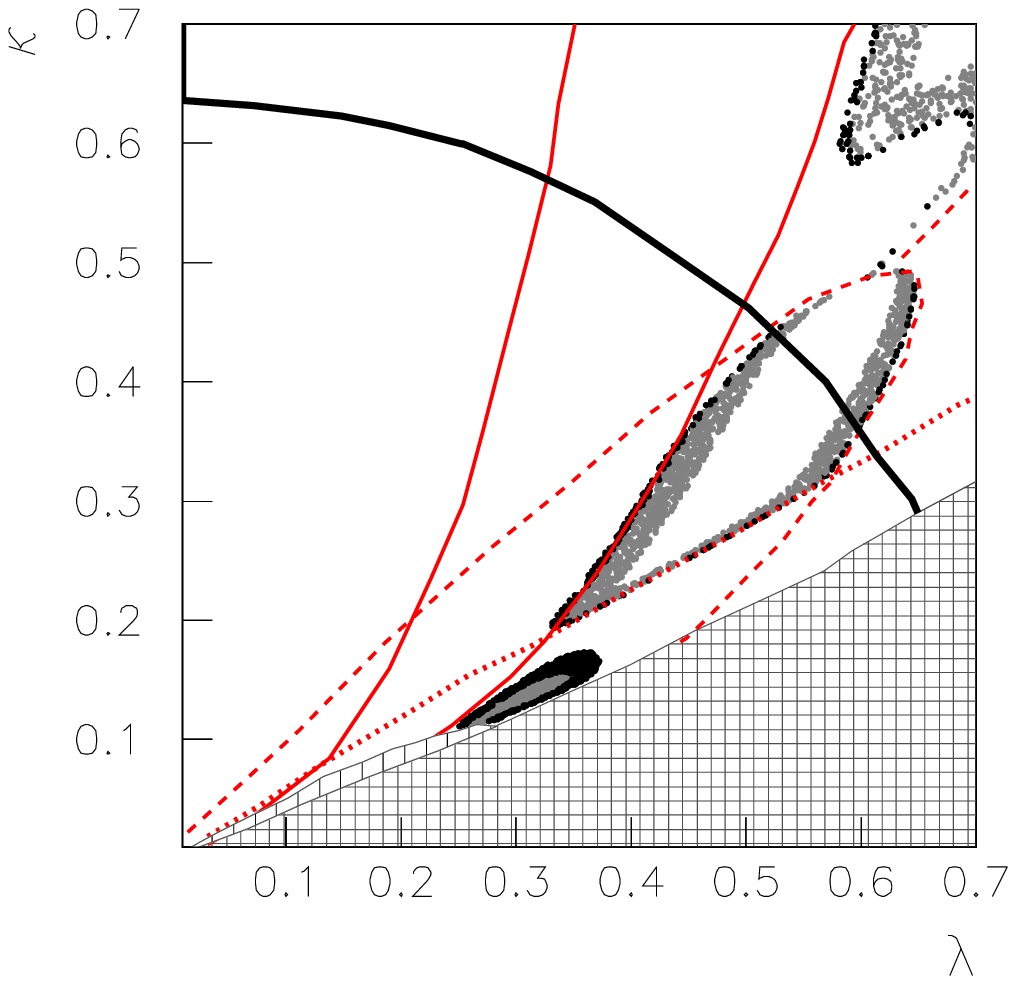,
    width=90mm}\hspace*{-10mm} 
  \epsfig{file=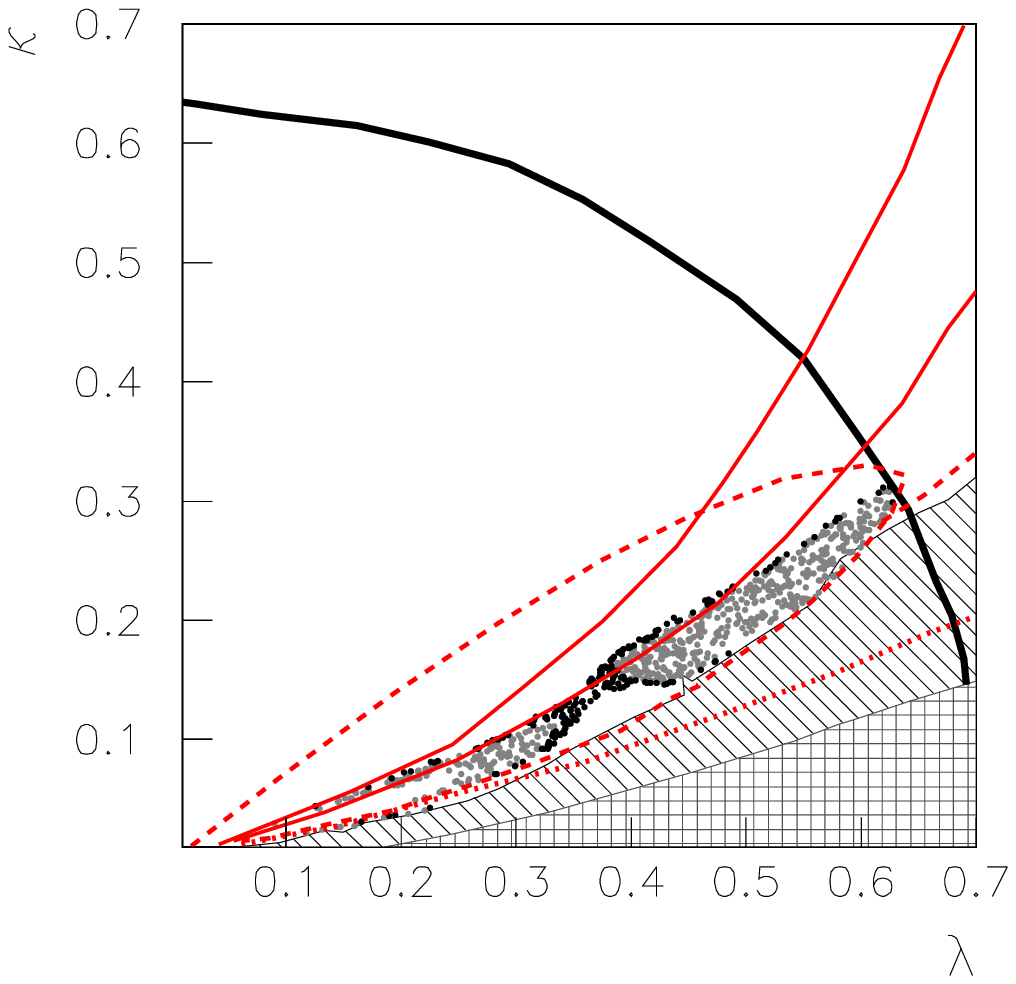, width=90mm}
  \vspace*{-1cm}
  \captions{
    $(\lambda,\kappa)$ parameter space with information about the
    neutralino relic density. On the left, an example with
    $M_1= 160$~GeV, $\tan\beta=5$, $A_\lambda=400$~GeV, 
    $A_\kappa=-200$~GeV, and $\mu=130$~GeV. 
    The gridded area is excluded due to the appearance of
    tachyons, while the vertically ruled area corresponds to the
    occurrence of unphysical minima. 
    The region above the thick
    black line is disfavoured because of the occurrence of a Landau
    pole below the GUT scale.
    The oblique ruled area is
    associated to those points that do not satisfy 
    LEP/Tevatron and/or BR($b\to s\,\gamma$) constraints, whereas 
    the bound on
    $\asusy$ is fulfilled in the whole plane. 
    The dark shaded (cyan) area corresponds to points
    which are experimentally viable, and whose relic density complies
    with the astrophysical bound of Eq.\,(\ref{om:mi}). Points in black
    are those in agreement with experimental constraints and 
    WMAP bounds (c.f. Eq.\,(\ref{om:wmap3})). 
    The dashed red lines indicate the resonances of the lightest
    neutralino annihilation channels through 
    the second lightest CP-even Higgs, $2\,\neumass=
    m_{h^{0}_2}$. In the region below the 
    red dotted line the 
    lightest neutralino mass is larger than the mass of the lightest
    Higgs. Along the red solid lines the
    neutralino mass is equal to the $Z$ and $W$ mass (from left to
    right, respectively). 
    On the right, the same example is shown, but with 
    $A_\kappa=0$ and $\mu=150$~GeV.
  }
  \label{fig:om:160} 
\end{figure}

We begin by taking $M_1=160$~GeV,
$A_\lambda=400$~GeV, $A_\kappa=-200$~GeV, and  $\mu=130$~GeV, with
$\tan \beta=5$, which according to Figs.\,\ref{g-2} and
\ref{fig:kl_160_200y400_150_-200_5}  
is consistent with the
bounds on $\asusy$ and BR($b \to s\,\gamma$), respectively. 
The results for the neutralino relic density are 
depicted in the $(\lambda,\kappa)$
plane on the left-hand side of Fig.\,\ref{fig:om:160}. On the
experimentally allowed area, grey dots stand for points which, 
in addition to experimental constraints, fulfil
$0.1\le\relic\le0.3$, whereas black dots represent points in agreement
with the WMAP constraint. 
Notice that in this case, no points are
excluded by LEP or Tevatron bounds.

For large values of $\kappa$ and small $\lambda$ (i.e., on the upper
left corner of the plots), the lightest neutralino is relatively heavy 
and has a mixed bino-Higgsino composition, since we have chosen 
$\mu\sim M_1$. Due to the large Higgsino component, the neutralino
relic density is very small, and cannot account for the
observed amount of dark matter.
As we move in the $(\lambda,\kappa)$ plane towards smaller values of
$\kappa$ and larger values of $\lambda$, 
the neutralino becomes lighter and has a
larger singlino component (in this example $N_{15}^2\lsim0.35$), 
and as a consequence, $\relic$ increases. 
As the neutralino mass decreases, some annihilation channels become
kinematically forbidden, such as annihilation into a
pair of $Z$ or $W$ bosons when $\neumass<M_Z$ or $\neumass<M_W$,
respectively. We have indicated these two thresholds in the figure
with red solid lines. Below these the resulting relic density can be 
large enough to fulfil the WMAP constraint. 
Variations in the Higgs sector also affect the calculation of the
neutralino abundance. On the one hand, the mass and composition of the
lightest Higgs also vary throughout the $(\lambda,\kappa)$ plane. 
Lighter
Higgses with a larger singlet composition are obtained for small
values of $\kappa$. In our case, $\neumass<m_{h_1^0}$
for large $\kappa$ and small $\lambda$, but
eventually, the Higgs becomes lighter and new annihilation channels
(the most important being $\neut\neut\to h_1^0h_1^0$ and $Zh_1^0$)
are available for the neutralino, thus decreasing its relic
density. 
The points where $\neumass=m_{h_1^0}$ are indicated with a
dotted red line in the plot. 
On the other hand, 
one also needs to take into account the existence of rapid
neutralino annihilation with the second-lightest CP-even Higgs, when  
$2\,\neumass=m_{h_2^0}$, which is responsible for
a further decrease in $\relic$. 
This is
indicated in the plot with a red dashed line.

As we can see, in the present example the correct relic density is
only obtained when either the singlino composition of the neutralino
is large enough or when the annihilation channels into $Z$, $W$, or
$h_1^0$ are kinematically forbidden. 
Interestingly, some allowed areas are very close to the
tachyonic border. The neutralino-nucleon cross section
can be very large in these regions, due to the presence of very light
singlet-like Higgses (in this example $S_{13}^2\approx0.99$).

The same example, but now with $A_\kappa=0$ and $\mu=150$ GeV
is shown on the right-hand side of
Fig.\,\ref{fig:om:160}. 
Once more, in order to reproduce the correct $\relic$ the neutralino
has to be either sufficiently light so that some annihilations
channels are 
closed or have a large singlino component. In this particular case the
singlino component of $\neut$ can be even larger, 
with $N_{15}^2\sim0.9$ in the allowed area with very low $\kappa$. 
Notice, however, that the region in the vicinity of the tachyonic area
is excluded by experimental bounds.

\begin{figure}[!t]
  \hspace*{-7mm}
  \epsfig{file=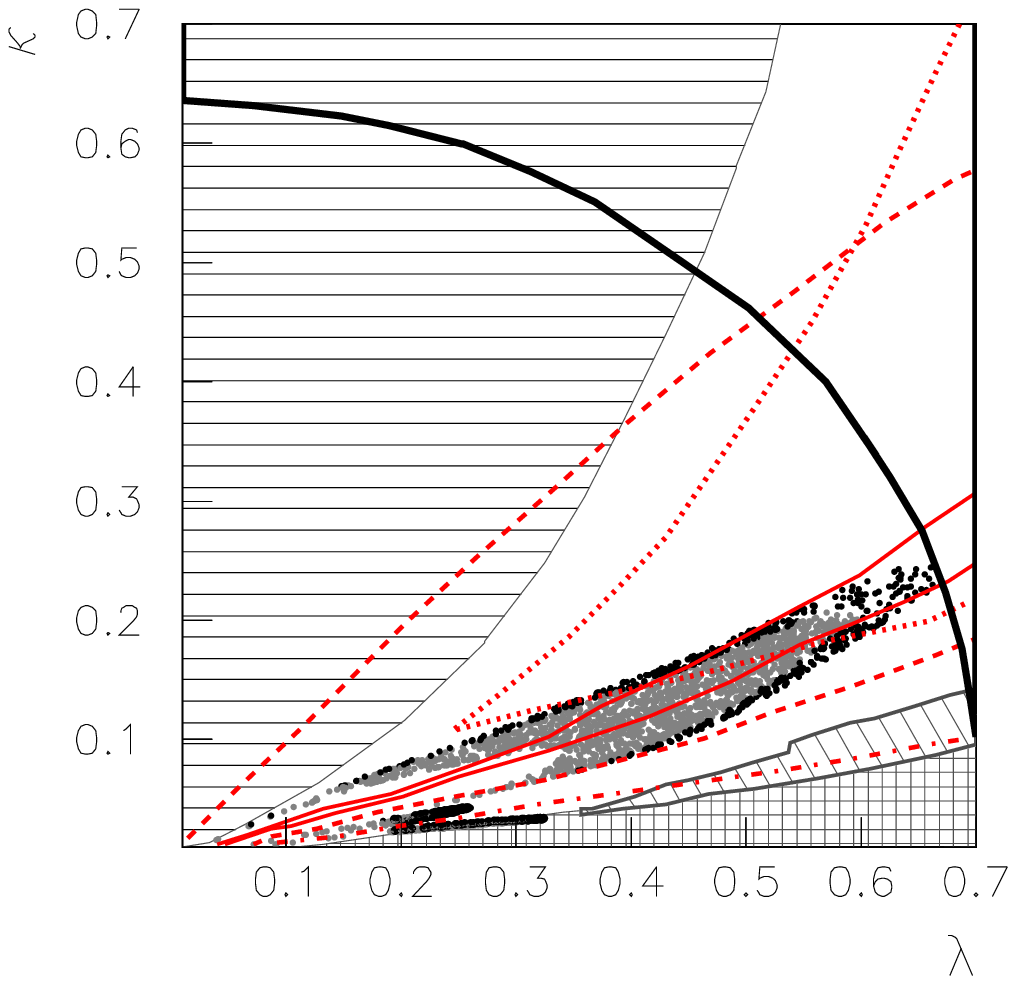,
    width=90mm}\hspace*{-10mm} 
  \epsfig{file=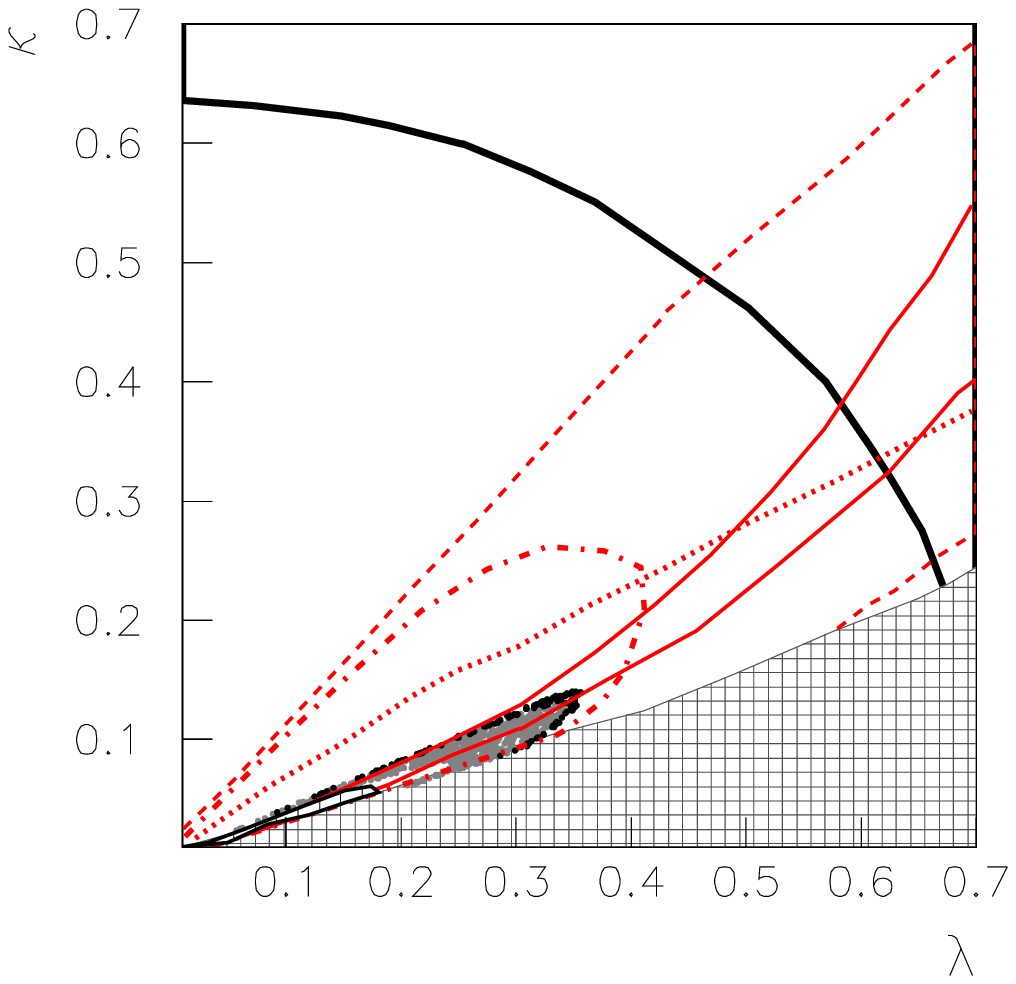,
  width=90mm} 
  \vspace*{-1cm}
  \captions{On the left, the same as in Fig.\,\ref{fig:om:160} but with 
    $M_1= 330$~GeV, $\tan \beta=5$, $A_\lambda=570$~GeV,  
    $A_\kappa=-60$~GeV, with $\mu=160$~GeV. The
    area to the right of the red dotted line has $\neumass<m_{h_1^0}$. 
    Along the red dot-dashed line, $2\,\neumass=M_{Z}$, leading to 
    rapid neutralino annihilation.
    On the right, the same with 
    $M_1= 500$~GeV, $\tan \beta=5$, $A_\lambda=400$~GeV,
    $A_\kappa=-150$~GeV, with $\mu=130$~GeV. 
    In this example the red dot-dashed line indicates resonances on
    the neutralino annihilation mediated by the lightest pseudoscalar
    when $2\,\neumass=m_{a_1^0}$. In both examples the resulting
    $\asusy$ is outside the experimental $2\sigma$ region.
  }
  \label{fig:om:330and500}
\end{figure}

In order to study the importance of the neutralino composition, 
we will now consider variations in the gaugino masses. To begin with, 
we increase the bino mass and take
$M_1=330$~GeV, thereby decreasing the bino component of the lightest
neutralino.  
Such an increase of the gaugino masses implies a reduced contribution
to the
muon
anomalous magnetic moment.
We obtain
$\asusy\approx7.2 \times 10^{-10}$ (see Fig.\,\ref{g-2}), 
more than $2\sigma$ away from the
central value and therefore disfavoured. 
The resulting $(\lambda,\kappa)$ plane is represented on the left-hand
side of Fig.\,\ref{fig:om:330and500}. 
Since the Higgsino component has increased with respect to the
previous examples, the resulting relic density for the neutralino in
the region with large $\kappa$ and small $\lambda$ is even smaller. 
Once more, in order to have the correct $\relic$ we need to go to
regions of the parameter space where some annihilation channels are
not kinematically allowed and/or the neutralino is
singlino-like. Notice also that the neutralino is in general 
heavier in this
example and therefore the lines with $\neumass=M_Z$ and $\neumass=M_W$ 
are shifted to lower values of $\kappa$. Also, the region with
$\neumass<m_{h_1^0}$ is modified and now corresponds to the area on the
right of the dotted red line. 
Finally, we must take into account the possible resonances along which
$\relic$ decreases. In this
example, rapid annihilation of neutralinos occurs via
CP-even Higgs exchange when $2\neumass=m_{h_1^0}$, which takes place
along the two upper red dashed lines. There is also a resonance with
the $Z$ boson when $2\neumass=M_{Z}$ which occurs along the lower red
dashed line.
It is worth noticing that, once more, 
a part of the region allowed by
experimental and astrophysical constraints lies close to the
tachyonic area, and could have a large $\crosssec$.

Finally on the right-hand side of Fig.\,\ref{fig:om:330and500} 
we show another example where the bino mass has been further
increased to $M_1=500$ GeV.
The contribution to the muon anomalous magnetic moment is also too
low. As shown in Fig.\,\ref{g-2}, $\asusy \approx4 \times 10^{-10}$, 
more than $2\sigma$ away from the
experimental value. Due to the further increase of the Higgsino
component and mass 
of the lightest neutralino, its relic density is even
smaller and compatibility with 
WMAP is only obtained when the neutralino is
lighter than the lightest Higgs and, at least, the $W$ boson. Notice
that in this example there is also a resonant annihilation
through the lightest CP-odd Higgs when $2\,\neumass\approx m_{a_1^0}$, 
which further decreases $\relic$. This constrains the allowed region
to small values of $\lambda$, in which the neutralino is mostly
singlino, $N_{15}^2\lsim0.8$.

To summarise, in these scenarios, neutralinos typically have a very
small relic density (insufficient to account for the dark matter in
the Universe) as a consequence of their large Higgsino
composition. Only when one moves towards regions of the parameter
space where the singlino composition is enhanced or the neutralino
mass is decreased (such that some annihilation channels become
forbidden) can the WMAP result be reproduced.

We have not yet addressed the other areas of the parameter space
(ii)-(iv). These other choices of signs for the different parameters
are associated to less favourable scenarios.
First, in cases (ii) and (iv) the experimental constraints rule out
the regions where $\neut$ has a large singlino
component. Therefore, the lightest neutralino is in general 
Higgsino-like
throughout all the allowed $(\lambda,\kappa)$ plane and, consequently,
its relic 
abundance is much below the favoured values, i.e., $\relic\ll0.1$. 
In case (iii), where large areas are excluded because of the
occurrence of tachyons, it is extremely complicated to find regions
that simultaneously fulfil the experimental and astrophysical
constraints. For this reason, in the following section we will limit
our analysis to case (i).

As a next step of our analysis, we will bring together all the
constraints so far explored, and after 
having ensured that we are indeed in
the presence of a viable NMSSM scenario (namely with the correct relic
density), we will investigate to which extent the lightest neutralino
can be detectable in dark matter experiments.

\subsection{Neutralino direct detection prospects}
\label{prospects:sigma}

After having discussed the new and the improved constraints on the
low-energy parameter space, we will now address whether or
not NMSSM neutralinos with a relic density in agreement with current
limits are likely to be detected by the present or the next generation
of dark matter detectors.

Although in our survey of the low-energy NMSSM parameter space we have
scanned over all combinations of signs (i)-(iv), as we already
mentioned, cases (ii)-(iv) present far less interesting situations
regarding the neutralino relic density. Even though 
one can find in the latter three
cases some challenging situations regarding direct detection prospects
\cite{Cerdeno:2004xw},
the new imposed constraints imply that finding  experimentally viable
areas, with a sizable $\sigma_{\tilde \chi^0_1 - p}$, becomes nearly
impossible. 
Thus, for the present study we will focus on case (i). We will go
through the
same examples as in the previous subsection.

\begin{figure}[!t]
  \hspace*{-7mm}
  \epsfig{file=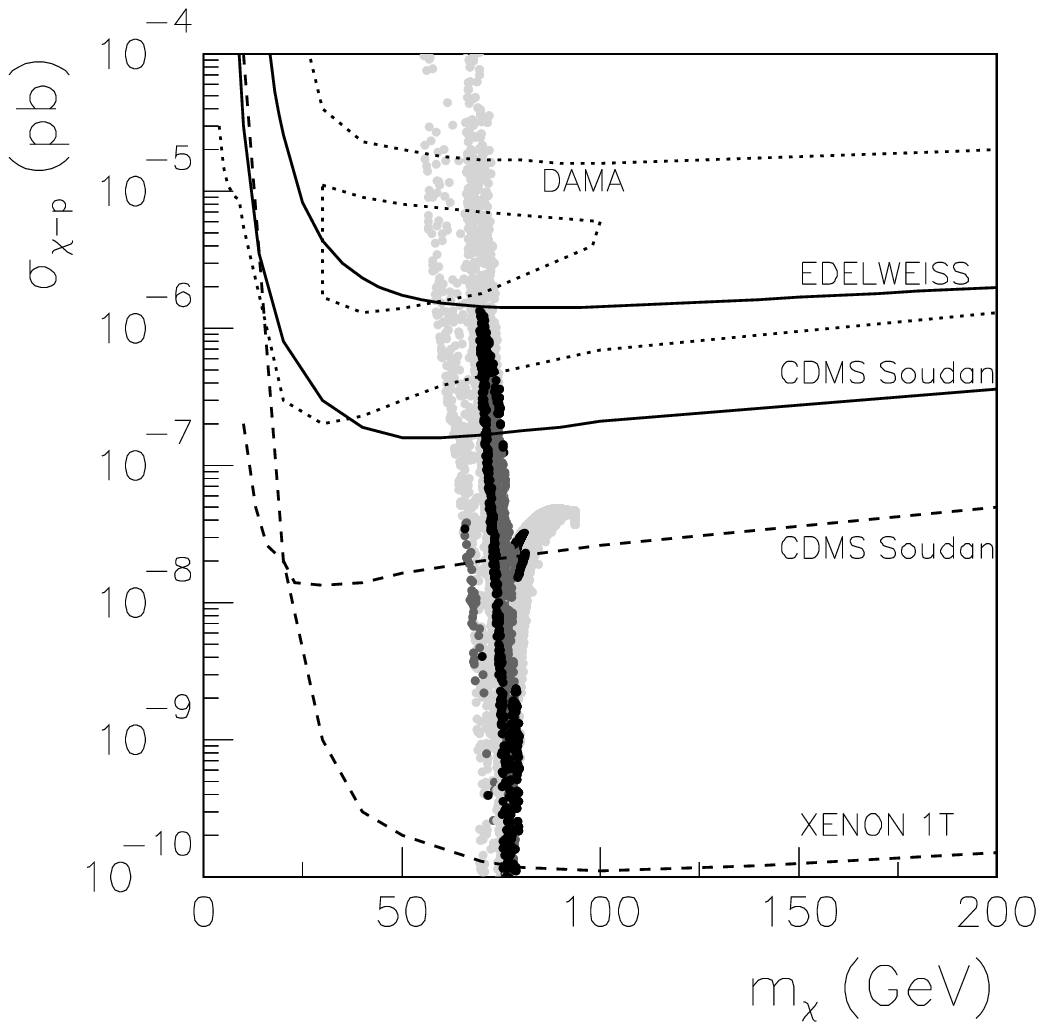,
    width=90mm}\hspace*{-10mm} 
  \epsfig{file=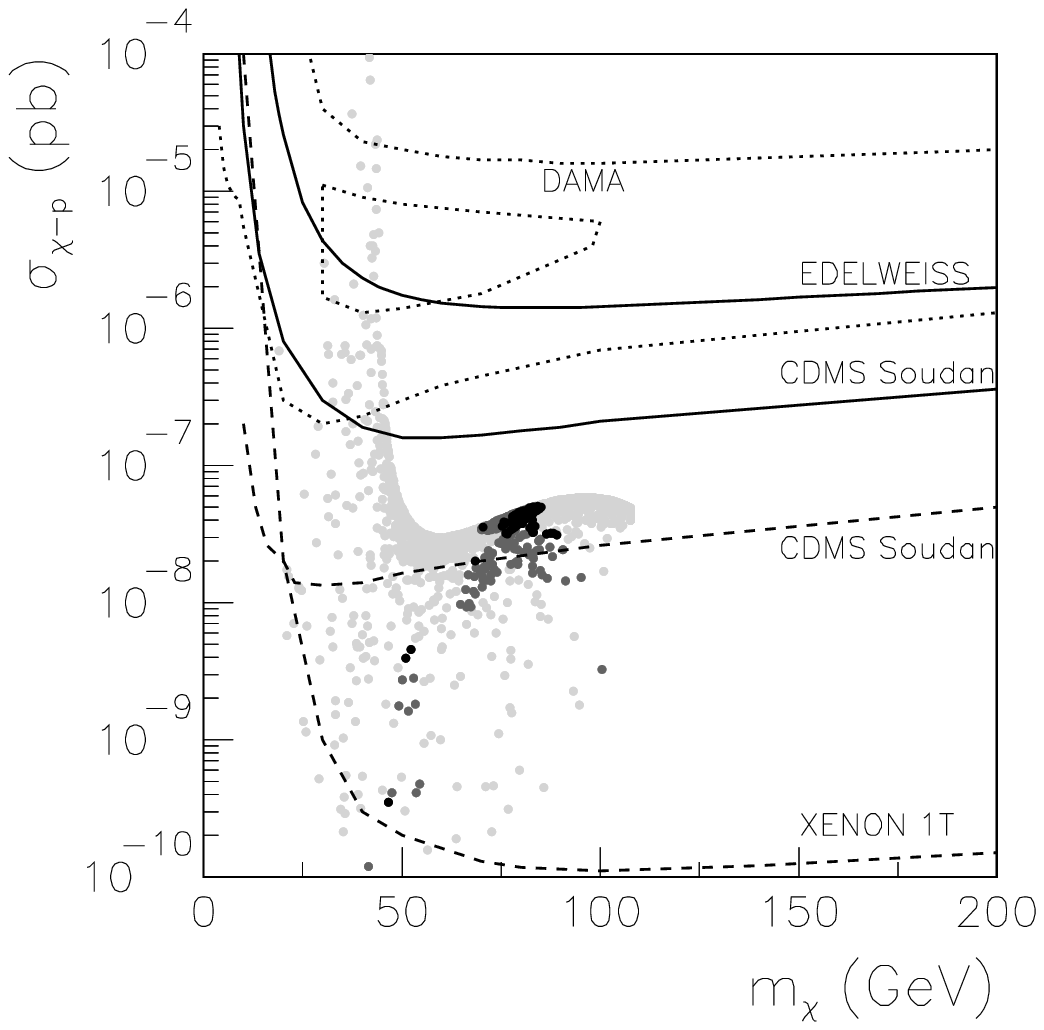, width=90mm}
  \vspace*{-1cm}
  \captions{ 
    Scatter plot of the scalar neutralino-nucleon cross section as a
    function of the lightest 
    neutralino mass. On the left,
    an example with $M_1= 160$ GeV, $\tan 
    \beta=5$,
    $A_\lambda=400$~GeV, $A_\kappa=-200$ GeV, and $\mu=130$ GeV. 
    All the points represented are in
    agreement with LEP/Tevatron, $\asusy$, and 
    BR($b\to s\,\gamma$) bounds. 
    Dark gray dots
    represent points which, in addition, fulfil $0.1\le\relic\le0.3$,
    whereas black dots are those in agreement with the WMAP constraint. 
    The sensitivities of
    present and projected  experiments are also depicted, with solid and
    dashed lines, respectively. 
    On the right we show 
    the same example but with $\mu=150$ GeV and $A_\kappa=0$. 
  }
  \label{fig:svhn:160}
\end{figure}

Let us start with the regime where $M_1 \approx \mu$, and as an
example ensuring compatibility with WMAP, choose
$A_\lambda=400$ GeV, $A_\kappa=-200$ GeV, $\mu=130$ GeV, with $\tan
\beta =5$ (corresponding to what was already depicted on the right
panel of Fig.\,\ref{fig:kl_160_200y400_150_-200_5} and the left of
Fig.\,\ref{fig:om:160}). 
As shown in the previous subsection, there exist regions in the
parameter space where the neutralino fulfils all experimental
constraints and has the correct relic density. The latter are
characterised by neutralinos with a significant singlino fraction
and/or a small mass. In this case, one of the allowed
regions is close to the tachyonic area and exhibits very light
singlet-like Higgses, potentially leading to large
detection cross sections.
This is indeed the case, as evidenced on the left-hand side of 
Fig.\,\ref{fig:svhn:160}, where the theoretical
predictions for $\crosssec$ are plotted versus the lightest 
neutralino mass.
The resulting $\crosssec$ spans several orders of magnitude, but,
remarkably, areas with $\crosssec\gsim 10^{-7}$ pb are found. These
correspond to the above mentioned regions of the parameter space with
very light 
singlet-like 
Higgses ($25\,{\rm GeV}\,\lsim m_{h_1^0}\lsim 50\,{\rm GeV}$ with
$S_{13}^2\gsim0.99$). 
The neutralino is a mixed singlino-Higgsino state with
$N_{15}^2\approx0.35$ and a mass around $75$ GeV.

The sensitivities of present and projected dark
matter experiments are also depicted 
for comparison. The small area bounded by dotted lines is
allowed by the DAMA experiment in the simple case of an isothermal
spherical halo model \cite{experimento1}. 
The larger area also bounded by dotted lines
represents 
the DAMA region when uncertainties to this simple model are taken into
account \cite{halo}. 
For the other experiments in the figure only the spherical halo has
been considered in their analyses. 
In particular, the (upper) areas bounded by solid lines are excluded
by EDELWEISS  \cite{edelweiss} 
\footnote{Since the exclusion area due to ZEPLIN I \cite{ZEPLINI} 
  is similar to
  EDELWEISS we have not depicted it here, nor in 
  any subsequent plot.} 
and CDMS Soudan \cite{soudan}. Finally, the dashed lines
represent the sensitivities of the projected  CDMS Soudan and XENON~1T 
\cite{xenon} experiments.

On the right-hand side of Fig.\,\ref{fig:svhn:160} we show the
resulting $\crosssec$ when the $\mu$ parameter and $A_\kappa$ are
varied to $\mu=150$ GeV, $A_\kappa=0$, for which the effect of the
different constraints on the $(\lambda,\kappa)$ plane were  
represented on the right-hand side of Fig.\,\ref{fig:om:160}. 
Since the areas of the parameter space with very light Higgses are
ruled out by experimental constraints the detection cross section is
not as large as in the previous examples. In the regions consistent
with both experimental and astrophysical 
constraints the lightest Higgs mass is in the range $80\,{\rm GeV}\,
\lsim m_{h_1^0}\lsim 
120\,{\rm GeV}$, thus
leading 
to $\crosssec\lsim5\times10^{-8}$~pb, within the sensitivity of
projected dark matter experiments, such as CDMS Soudan.

Let us now investigate the effect of changing the neutralino
composition by modifying the bino mass. As commented in
\cite{Cerdeno:2004xw}, the largest values of the neutralino detection 
cross section were obtained for a mixed singlino-Higgsino composition, 
when $\mu\lsim M_1<M_2$.
In order to enhance the Higgsino composition we will consider examples
where $M_1$ is increased with respect to the $\mu$-parameter.
Such neutralinos annihilate more efficiently, thus
leading to a reduced $\relic$, so that the astrophysical constraint
becomes more stringent. 
Nevertheless, as seen in the previous subsection,
it is still possible to find areas of the parameter space with the
correct relic density while simultaneously fulfilling all experimental
constraints. These regions corresponded to light singlet-like Higgses,
which can potentially lead to sizable detection cross sections.

\begin{figure}[t]
  \hspace*{-7mm}
  \epsfig{file=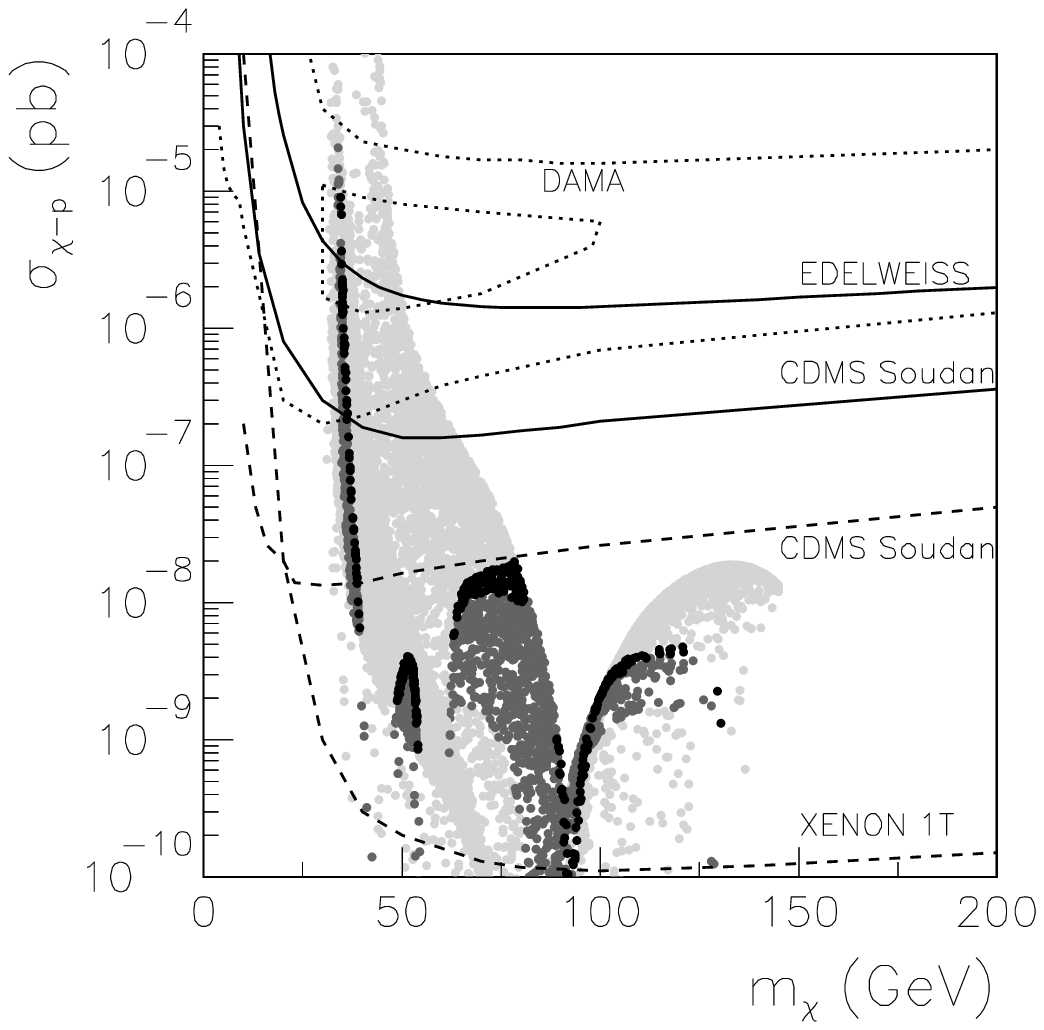,
    width=90mm}\hspace*{-10mm}
  \epsfig{file=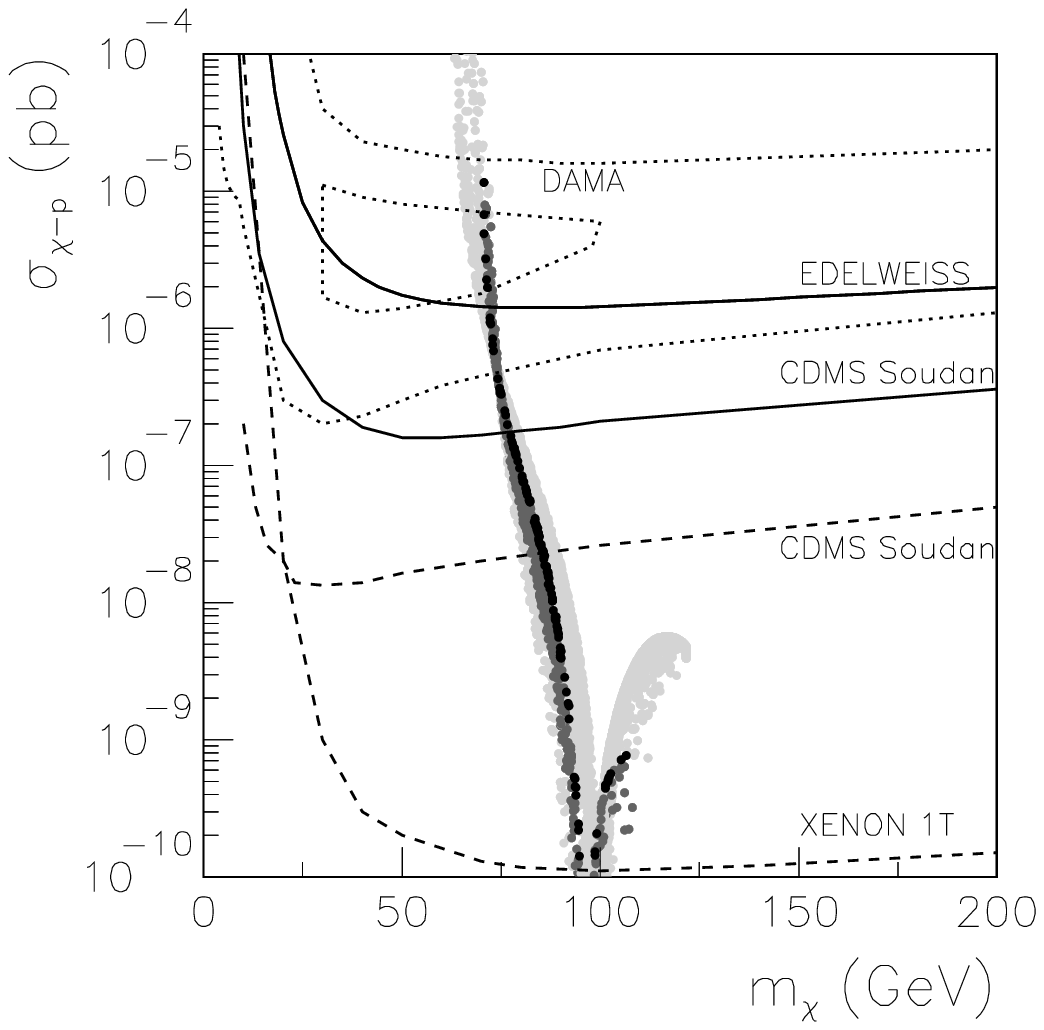,
    width=90mm}
  \vspace*{-1cm}
  \captions{The same as in Fig.\,\ref{fig:svhn:160} but for an example
  with $M_1= 330$ GeV, $\tan
  \beta =5$, $A_\lambda=570$   GeV, $A_\kappa=-60$ GeV, with
  $\mu=160$ GeV (left) and $M_1= 500$ GeV, 
  $\tan \beta=5$ 
  $A_\lambda=400$   GeV, $A_\kappa=-150$ GeV, with $\mu=130$ GeV
  (right). In both examples, the resulting
  $\asusy$ is outside the experimental $2\sigma$ region.
  }
  \label{fig:svhn:330y500}
\end{figure}

First, $M_1=330$ GeV will be taken, for an example with $\mu=160$ GeV,
$A_\lambda=570$   GeV, $A_\kappa=-60$ GeV, and $\tan\beta =5$. The
parameter space for this case was represented in
Fig.\,\ref{fig:om:330and500}, were we showed the effect of resonant
annihilation channels on the allowed regions. 
The theoretical predictions for neutralino direct detection are shown
in Fig.\,\ref{fig:svhn:330y500}.
In this plot, the various resonances appear as funnels in the
predicted $\crosssec$ for the regions
with the correct $\relic$ at the corresponding values of the 
neutralino mass ($\neumass\approx M_Z/2$ and $\neumass\approx
m_{h_1^0}/2$). 
Below the resonance with the $Z$ boson, light neutralinos are obtained
$\neumass\lsim M_Z/2$ with a large singlino composition which have
the correct relic abundance. The lightest Higgs is also singlet-like
and very light, leading to a very large detection cross section,
$\crosssec\gsim10^{-6}$ pb. This corresponds to the allowed area of
the $(\lambda,\kappa)$ plane which lies in the vicinity of the
tachyonic region in Fig.\,\ref{fig:om:330and500}.

Remember however that these two 
examples with a larger bino mass 
were disfavoured by the resulting muon
anomalous magnetic moment, as
it was illustrated in Fig.\,\ref{g-2}.

One more example, this time for $M_1=500$ GeV,
$\mu=130$ 
GeV, $A_\lambda=400$ GeV, $A_\kappa=-150$ GeV, and $\tan \beta =5$ is
represented in Fig.\,\ref{fig:svhn:330y500} and
shows how large detection cross sections can also be achieved for
heavier neutralinos. In this case (whose parameter space was
illustrated and discussed in Fig.\,\ref{fig:om:330and500}) neutralino
detection cross sections as large as $\crosssec\approx10^{-5}$ pb are
possible while fulfilling experimental and astrophysical
constraints. Once more, the occurrence of light singlet-like Higgses
is crucial for enhancing $\crosssec$ and the sizable singlino
component of the lightest neutralino ($N_{15}^2\approx0.9$) reduces
the annihilation cross section and ensures the correct relic density.

\begin{figure}[t]
  \hspace*{-7mm}
  \epsfig{file=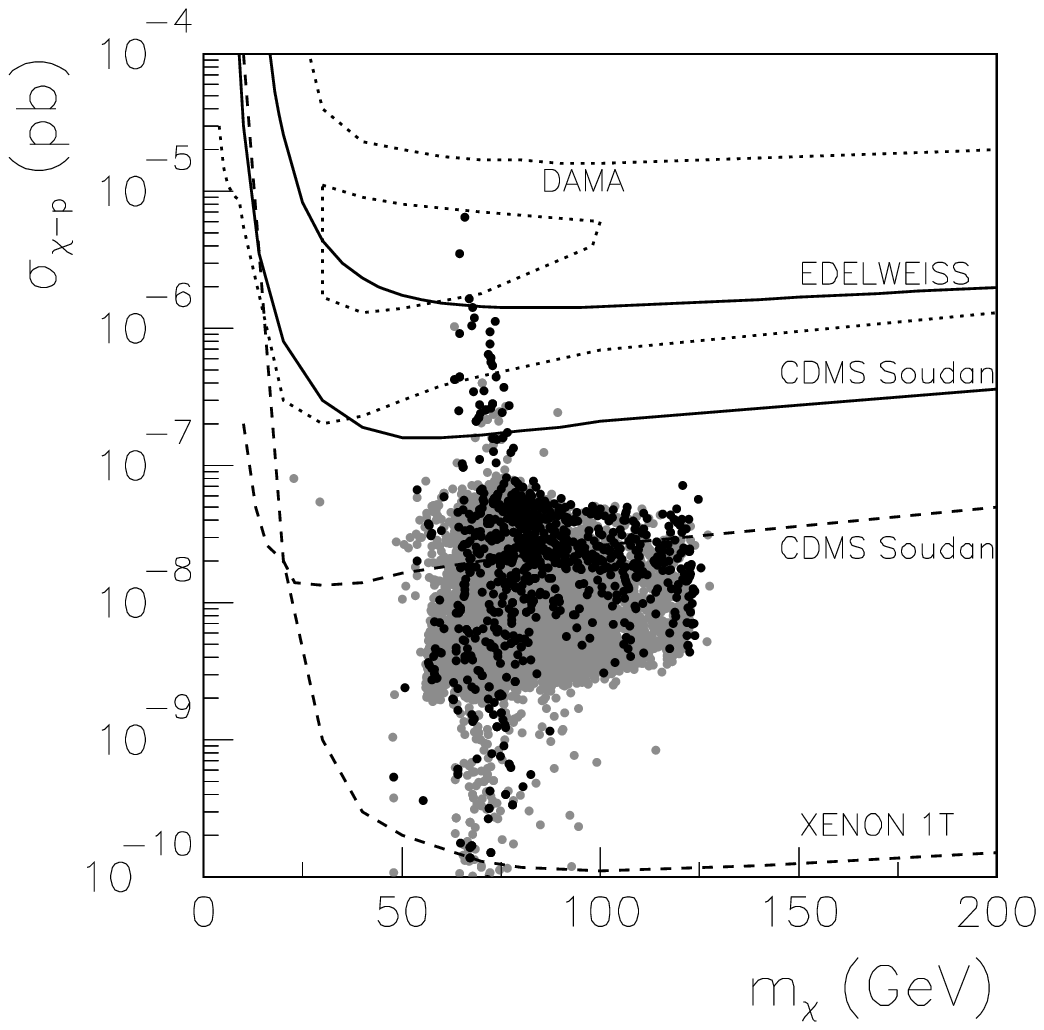, width=90mm}
  \hspace*{-10mm}
  \epsfig{file=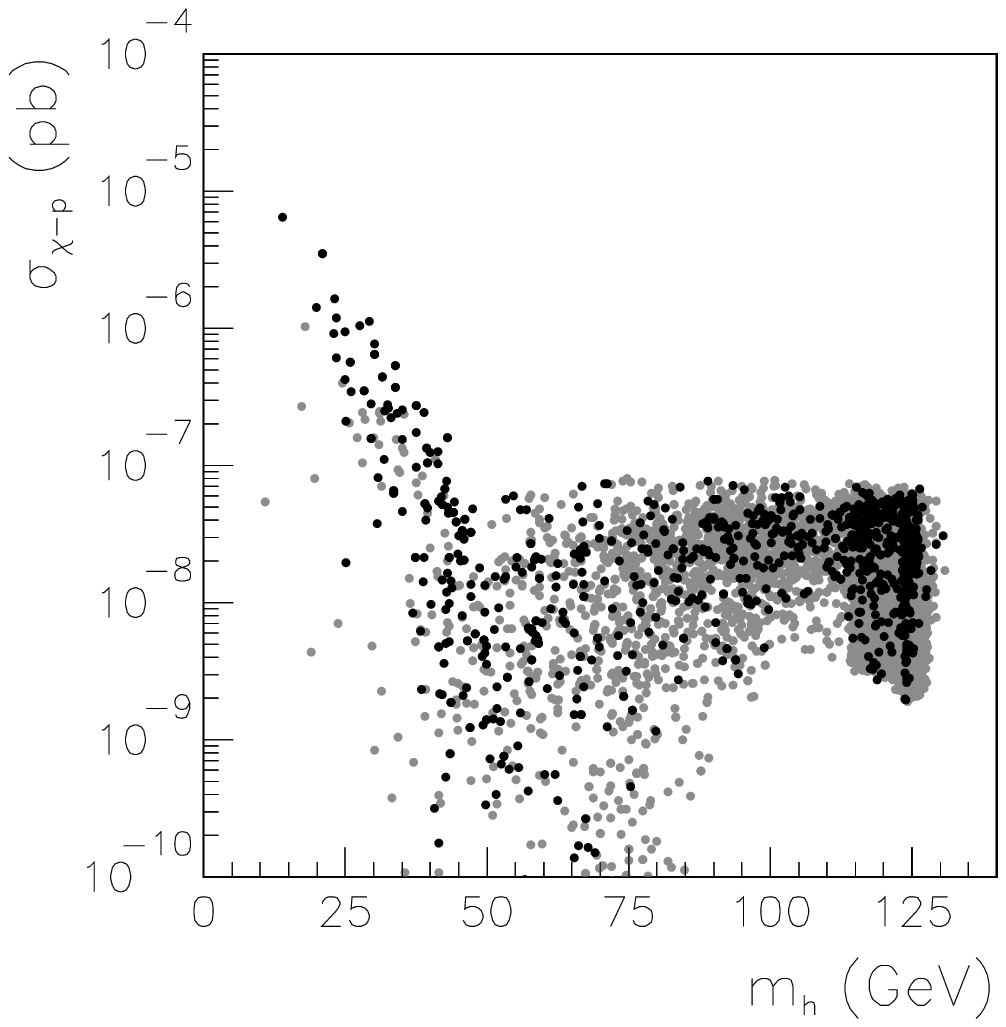, width=90mm}
  \vspace*{-1cm}
  \captions{
    Scatter plot of the 
    neutralino-nucleon cross section as a function of the neutralino
    mass (left) and as a function of the lightest CP-even Higgs mass
    (right) for an example with
    $\tan \beta=5$, and the remaining parameters in the ranges 
    $0.01\le\lambda,\kappa\le0.7$, 
    $110\ {\rm GeV} \lesssim M_2 \lesssim 430\ {\rm GeV}$, 
    $-300\ {\rm GeV}\lesssim A_\kappa \lesssim 300\ {\rm GeV}$,
    $-800\ {\rm GeV} \lesssim A_\lambda \lesssim 800\ {\rm GeV}$, and 
    $110\ {\rm GeV}<\mu<300\ {\rm GeV}$.
    All the points represented are in
    agreement with LEP/Tevatron, $\asusy$, and 
    BR($b\to s\,\gamma$) constraints, and have a relic density
    in agreement with the astrophysical bound (grey dots) or the WMAP
    constraint (black dots).
  }
  \label{summary1}
\end{figure}

Finally, we show in Fig.\,\ref{summary1} a scatter plot of the
theoretical predictions for the neutralino-nucleon cross section as a
function of the neutralino mass and the lightest Higgs mass 
when $\lambda$, $\kappa$, $M_1$, the $\mu$
parameter and the trilinear terms, $A_\lambda$ and $A_\kappa$ are
varied while keeping $\tan\beta=5$. 
In order to satisfy the $\asusy$ constraint, a small slepton mass,
$m_{L,E}=150$~GeV, has been used. 
Only the points which are
in agreement with LEP/Tevatron, BR($b\to s\,\gamma$), and $\asusy$
limits  
and which, in addition, are consistent with the astrophysical bound
or the WMAP constraint on the relic density
are plotted. 
We clearly see how large detection cross sections are correlated to
the presence of very light Higgses ($m_{h_1^0}\lsim50$ GeV).
Neutralinos fulfilling all constraints and within the reach of dark
matter experiments are possible with $50\,{\rm GeV}\lsim \neumass\lsim
130\,{\rm GeV}$. 
The upper bound on the neutralino mass is due to the lightest 
stau becoming the
LSP. If the slepton mass is increased, heavier neutralinos can be
found but the resulting $\asusy$ 
is soon outside the experimentally allowed range.

\section{Conclusions}
\label{conclusions}

We have extended the systematic analysis started in
\cite{Cerdeno:2004xw} of the low-energy parameter space of the
Next-to-Minimal Supersymmetric Standard Model (NMSSM), studying the
implications of experimental and astrophysical constraints 
on the direct detection of neutralino dark matter. We
have computed the theoretical predictions for the scalar
neutralino-proton cross section, $\crosssec$, 
and compared it with the
sensitivities of present and projected dark matter experiments. 
In the computation we have taken into account all available
experimental bounds from LEP and Tevatron, including
constraints coming from $B$ and $K$ physics, as well as the
supersymmetric    
contribution to the muon anomalous magnetic moment, $\asusy$. 
Finally, the relic abundance of neutralinos has also been 
computed and consistency 
with astrophysical constraints imposed.

We have found very stringent constraints on the parameter space coming
from low-energy observables.
On the one hand,
$\asusy$ is generally very small unless very light slepton
($m_{L,E}\lsim200$ GeV) and gaugino masses ($M_1\lsim210$ GeV)
are considered, and slepton trilinear couplings modified
in order to increase the $LR$ mixing in the smuon mass matrix. 
On the other hand, the contribution to BR($b \to s
\gamma$) is sizable, mostly due to the smallness of the charged Higgs
mass, so that regions with $\tan\beta\lsim3$ are disfavoured.

Regarding the neutralino relic density, regions of the parameter space
can be found where $\relic$ is in agreement with the WMAP
constraint. This is possible when either the neutralino mass is small
enough for some annihilation channels to be kinematically forbidden or
when the singlino component of the lightest neutralino is large enough
to suppress neutralino annihilation.

Remarkably, some of the regions fulfilling all the experimental and
astrophysical 
constraints
display very light Higgses, $m_{h_1^0} \sim 50 $~GeV, which are
singlet-like, $S_{13}^2\gsim0.9$, thus allowing a sizable
increase of the neutralino-nucleon cross section. 
Neutralinos with a detection cross-section 
within the reach of dark matter
experiments are therefore possible, and have a mass
in the range $50\,{\rm GeV}\lsim \neumass\lsim
130\,{\rm GeV}$. These neutralinos have a mixed singlino-Higgsino
composition and are therefore characteristic of the NMSSM.

\section*{Acknowledgements}
\label{ack}

D.~G. Cerde\~no is supported by the program ``Juan de la Cierva'' of
the Ministerio de Educaci\'on y Ciencia of Spain.
The work of E. Gabrielli was supported by
the Academy of Finland (Project number 104368). 
The work of C. Mu\~noz was supported
in part by the Spanish DGI of the
MEC under Proyecto Nacional FPA2006-05423,
by the European Union under the RTN program
MRTN-CT-2004-503369, and under the ENTApP Network of the ILIAS project 
RII3-CT-2004-506222.
Likewise, the work of D.~G. Cerde\~no, D.~E. L\'opez-Fogliani, and
C. Mu\~noz, was also supported in part by the Spanish DGI of the
MEC under Proyecto Nacional FPA2006-01105
and under Acci\'on Integrada Hispano-Francesa HF-2005-0005, and
by the Comunidad de Madrid under Proyecto HEPHACOS, Ayudas de I+D
S-0505/ESP-0346. 
The work of A.~M.~Teixeira was supported by the French ANR
project PHYS@COL\&COS.

\end{document}